\pdfoutput=1
\documentclass[3p]{elsarticle}
\usepackage{lineno}
\usepackage{amssymb,amsbsy,nicefrac}
\usepackage{color}
\usepackage{graphics}
\usepackage{graphicx}
\modulolinenumbers[1]

\journal{Comput. Methods Appl. Mech. Engrg.}

\makeatletter
\@addtoreset{equation}{section}

\makeatother

\biboptions{sort&compress}

\begin{document}

\begin{frontmatter}

\title{Improvement of accuracy of the spectral element method for elastic wave
computation using modified numerical integration operators}

\author[as]{Kei Hasegawa\corref{cor1}}
\ead{khase@earth.sinica.edu.tw}
\author[ipgp]{Nobuaki Fuji}
\author[as]{Kensuke Konishi}


\cortext[cor1]{Corresponding author}
\address[as]{Institute of Earth Sciences, Academia Sinica, 128 Academia Road Sec. 2, Nangang,\\ Taipei 11529, Taiwan}
\address[ipgp]{Institut de Physique du Globe de Paris, 1 rue Jussieu, 75238 Paris Cedex 05, France}

\begin{abstract}
We introduce new numerical integration operators which compose the mass and
stiffness matrices of a modified spectral element method for simulation of elastic wave propagation.
While these operators use the same quadrature nodes as does the original spectral element method, they are designed in order that their lower-order contributions to the numerical dispersion error cancel each other.
As a result, the modified spectral element method yields two extra-orders of accuracy,
and is comparable to the original method of one order higher.
The theoretical results are confirmed by numerical dispersion analysis and examples of computation
of waveforms using our operators.
Replacing the ordinary operators by those proposed in this study could be a non-expensive solution to improve the accuracy.
\end{abstract}


\end{frontmatter}

\section{Introduction}\label{sec.intro}
Finite element methods (FEMs) for computation of the elastic wave equation
have greatly contributed to seismology and earthquake engineering
\cite{PCR09, GKO13,FI16}.
Notably,
the spectral element method (SEM) is most widely used
in the past twenty years \cite{KV98,KT99}.
For elastic wave computation,
the SEM is usually associated with
the Gauss--Lobatto--Legendre (GLL) quadrature rule and Lagrange polynomial basis
defined on hexahedral elements,
because this choice
leads to an explicit time-marching scheme without loss of accuracy of computation.
Detailed descriptions are available on \cite{KS05,I16}.

In applications of FEMs to elastic wave computation in complex underground structures,
there may still exist difficulty concerning grid-generation.
According to dispersion and stability analyses \cite{CFL28,NR50,MB82,BS07,SO08a,SO08b,OS10},
it is preferable to use almost the same number
of grid points per wavelength throughout the medium:
i.e., in terms of accuracy,
the number of grid points per wavelength
should be sufficiently large to suppress numerical dispersion \cite{MB82,BS07,SO08a,SO08b,OS10};
conversely,
an unnecessarily large number of grid points (or small grid intervals) may increase
the total number of time steps
as well as computational cost required for each time step,
since time intervals
should be much
smaller than the time for a wave train to pass through one grid interval \cite{CFL28,NR50,BS07}.
In other words, we need a dense grid for a region of a lower propagation velocity,
and a coarse grid for a higher one,
since the length of a wavelet depends on the propagation velocities.
However, this condition makes the grid-generation more complicated
as velocity structures become complex.
Instead of regulating the number of grid points per wavelength,
a regional increase/decrease of the order of elements would effectively improve the accuracy and efficiency.
However, for the Legendre-type SEM
(hereafter simply called the SEM),
 in particular,
the non-equispaced distribution of the GLL nodes makes
it difficult to connect elements of heterogeneous orders,
without rather complicated implementations \cite{BNM90}.
Moreover,
a use of higher-order elements can degrade flexibility
to match the grid geometry with a model structure,
compared with low-order elements with the same number of grid points.
As another disadvantage,
the time interval should be chosen to be smaller
as the order of elements is higher,
since GLL nodes tend to concentrate toward the edges of each element
and then the smallest distance between adjacent nodes becomes smaller \cite{I16}.
Therefore, a superconvergent SEM which can improve the accuracy
without increasing the number and order of elements
is highly desired in order to extend SEM applications.

In this paper,
we introduce modified numerical integration operators for the SEM for elastic wave computation.
While our operators are associated with the GLL nodes as those of the SEM, they are designed to yield higher accuracy of computation
without increasing the number and order of elements. 
As an alternative to the use of a dense grid or higher-order elements
to improve the accuracy of the SEM computation,
we suggest that replacing the ordinary SEM operators by our operators throughout all elements,
or embedding our operators for a region of a lower propagation velocity,
surrounded by the ordinary SEM operators
for regions of adequate accuracy depending on the demand of a user,
without making any change in grid settings.

Studies for superconvergent SEM/FEMs
are traceable back to the following previous works.
Marfurt~\cite{M84} suppressed the numerical dispersion of the linear FEM by
blending the consistent and lumped mass matrices in an empirical way.
Seriani and Oliveira~\cite{SO07} extended his study for the SEM.
Ainsworth and Wajid~\cite{AW10} analytically found the
optimal value of the blending ratio such that
the numerical dispersion of the SEM is minimized.
Note also that their ideas have been applied to the isogeometric analysis method,
which is well-suitable for structures with smooth curved surfaces \cite{CDP17,I17,PDC17,ID17,DBPC18}.
While the above studies are for computation of the Helmholtz equation,
an extension to the Maxwell equations is given by \cite{WA12}.
In the field of computational seismology,
a study for the elastic wave equation is given by \cite{GT95}.
They introduced a general criterion to minimize modal errors based on a perturbation approach, and gave
a superconvergent linear FEM for computing elastic wave propagation.
In this paper, we begin with a review of the criterion given by \cite{GT95},
and thereby give modified numerical integration operators which minimize
the numerical dispersion of the SEM for elastic wave computation.
Note that our results have some parts related to the optimally blending integration operator previously introduced by \cite{AW10}, which are also derived in this paper in a simpler way.
Further, we show new results for elastic wave computation.

\section{Review: general analysis for modal error estimation}\label{sec.rev95}
While the notations are different from those used in \cite{GT95},
the results in this part have been previously given by \cite{GT95}.
The elastic wave equation with the free surface boundary condition
is
\begin{equation}
	\left\{
	\begin{array}{ll}
	\displaystyle{
	\rho \partial_{tt} u_\alpha
	=\sum_{\beta,\xi,\eta =x,y,z}
	\partial_\beta
	\left(c_{\alpha\beta\xi\eta} \partial_\eta u_\xi\right)
	}
	+f_\alpha &\mbox{in}\hspace{0.5em}V
	\\   \vspace{-0.8em}\\
		\displaystyle{\sum_{\beta,\xi,\eta= x,y,z}
		n_\beta  \left(c_{\alpha\beta\xi\eta} \partial_\eta u_\xi \right) =0}
	& \mbox{on}\hspace{0.5em}\partial V,
	\end{array}
	\right.
		   \label{3r2}
\end{equation}
where
$V$ and $\partial V$ denote the volume of the medium and
its surface boundary,
Greek indices $\alpha$, $\beta$, $\xi$, and $\eta$ represent dummy indices for
$x$, $y$, and $z$-axes,
$u_\alpha(t,\vec{x})$ is the $\alpha$-component of the displacement
at the time $t$ and position $\vec{x}=(x,y,z)$,
$\partial_\alpha$ denotes partial differentiation with respect to the $\alpha$-axis,
$\partial_{tt}$ denotes the second-order temporal differentiation,
$\rho(\vec{x})$ is the density,
$c_{\alpha\beta \xi \eta} (\vec{x})$ are the elastic moduli,
$f_\alpha(t,\vec{x})$ is the external body force,
and $n_\alpha(\vec{x})$ is the $\alpha$-component of the unit vector normal to $\partial V$.
In this paper, we focus on isotropic cases.
For these cases, the elastic moduli are given by
\begin{equation}
	c_{\alpha\beta\xi\eta} =
	\lambda \delta_{\alpha\beta}\delta_{\xi\eta} + \mu (\delta_{\alpha \xi} \delta_{\beta\eta}+
	\delta_{\alpha\eta}\delta_{\beta\xi} ),
		   \label{iso_elastic_moduli}
\end{equation}
where $\lambda(\vec{x})$ and $\mu(\vec{x})$ are the Lam\'{e} parameters,
and $\delta_{\alpha\beta}$ is the Kronecker-delta.
Eq.~(\ref{3r2}) with the isotropic medium (\ref{iso_elastic_moduli}) is converted to the following variational form of the elastic wave equation:
\begin{equation}
   \partial_{tt} M(\vec{w}, \vec{u})=-K(\vec{w}, \vec{u}) +F(\vec{w}),
   \label{4r2}
\end{equation}
where
$\vec{w}=(w_x, w_y, w_z)$ is the weight vector function,
$\vec{u}=(u_x, u_y, u_z)$ is the displacement,
and $M$, $K$, and $F$ are
\begin{eqnarray}
   M(\vec{w},\vec{v}) &=& \sum_{\alpha=x,y,z} \int_V w_\alpha \rho v_\alpha dV
   \label{exact2}\\
   K(\vec{w}, \vec{v})&=&
   \sum_{\alpha,\beta =x,y,z} \int_V
   \left[ (\partial_\alpha w_\alpha)\lambda (\partial_\beta v_\beta)
    + (\partial_\beta w_\alpha)\mu (\partial_\alpha v_\beta+\partial_\beta v_\alpha) \right]dV
   \label{exact3}\\
   F(\vec{w}) &=& \sum_{\alpha=x,y,z}\int_V w_\alpha f_\alpha dV
   \label{exact4}
\end{eqnarray}
with vectors $\vec{w}=(w_x, w_y, w_z)$ and $\vec{v}=(v_x, v_y, v_z)$.

The semi-discrete form of Eq.~(\ref{4r2})
may be formally written as follows:
\begin{equation}
   \partial_{tt} M^{num}(\vec{w}, \vec{u}^{\,num})=-K^{num}(\vec{w}, \vec{u}^{\,num}) +F^{num}(\vec{w}),
   \label{1}
\end{equation}
where $\vec{u}^{num}=(u_x^{num}, u_y^{num}, u_z^{num})$
is the numerical solution.
Numerical operators $M^{num}$, $K^{num}$, and $F^{num}$ are given by
\begin{eqnarray}
   M^{num}(\vec{w},\vec{v}) &=& \sum_{\alpha=x,y,z}\mbox{n.i.}\int_V w_\alpha \rho v_\alpha dV
   \label{2}\\
   K^{num}(\vec{w}, \vec{v})&=&
   \sum_{\alpha,\beta=x,y,z}\mbox{n.i.}\int_V
   \left[ (\partial_\alpha w_\alpha)\lambda (\partial_\beta v_\beta)
    + (\partial_\beta w_\alpha)\mu (\partial_\alpha v_\beta+\partial_\beta v_\alpha) \right]dV
    \label{3}\\
   F^{num}(\vec{w})&=&\sum_{\alpha=x,y,z}\mbox{n.i.}\int_V w_\alpha f_\alpha dV,
   \label{4}
\end{eqnarray}
where ``n.i.'' is an abbreviation for ``numerical integration''
by which the integral is approximated according to a numerical integration rule.

Herein, we consider time-harmonic normal mode oscillations
with no external force ($f_\alpha=0$):
\begin{equation}
	\vec{u}(t,\vec{x}) = \vec{\Theta}_m(\vec{x}) \mathrm{e}^{-\mathrm{i}\Omega_m t}.
	\label{normalmode}
\end{equation}
where
$\mathrm{i}$ is the imaginary unit,
$\mathrm{e}$ is the Napier's constant,
$m$ represents the mode number,
and 
$\Omega_m$ and $\vec{\Theta}_m$  denote the 
eigenfrequency and eigenfunction of the $m$th-mode.
Substituting Eq.~(\ref{normalmode}) into Eq.~(\ref{4r2}) with $f_\alpha=0$,
the normal modes satisfy
the following eigenvalue problem:
\begin{equation}
  K(\vec{w}, \vec{\Theta}_m )=
  \Omega_m^2 M(\vec{w}, \vec{\Theta}_m).
   \label{ext_eigen}
\end{equation}
The  eigenfunctions are orthonormalized as follows:
\begin{equation}
   M( \vec{\Theta}_{m'}^*, \vec{\Theta}_m )
   = \delta_{m'm},
   \label{ext_orth}
\end{equation}
where asterisk indicates complex conjugate.

Similarly, numerically computed normal modes satisfy 
the following eigenvalue problem:
\begin{equation}
   K^{num}(\vec{w}, \vec{\Theta}_m^{num}) =
   (\Omega_m^{num})^2 M^{num}(\vec{w}, \vec{\Theta}_m^{num}),
   \label{num_eigen}
\end{equation}
where $\Omega_m^{num}$ and $\vec{\Theta}_m^{num}$ are the numerical eigenfrequency and eigenfunction of the $m$th-mode, respectively.
The numerical eigenfunctions are orthonormalized as follows:
\begin{equation}
   M^{num}([\vec{\Theta}_{m'}^{num} ]^*, \vec{\Theta}_m^{num})
   =\delta_{m'm}.
   \label{num_orth}
\end{equation}

We formally denote
the error of the numerical operators by $\delta M$ and $\delta K$,
and the error of the numerical eigenfrequency and eigenfunction of the
$m$th-mode
by $\delta \Omega_m$ and $\delta \vec{\Theta}_m$,
where
\begin{eqnarray}
   M^{num}(\vec{w},\vec{v})
   &=&M(\vec{w},\vec{v})+\delta M(\vec{w},\vec{v})
   \label{del_m}\\
      K^{num}(\vec{w},\vec{v})
   &=&K(\vec{w},\vec{v})+\delta K(\vec{w},\vec{v})
   \label{del_k}\\
   \Omega_m^{num} &=& \Omega_m +\delta \Omega_m\label{del_omega}\\
   \vec{\Theta}_m^{num} &=& \vec{\Theta}_m +\delta \vec{\Theta}_m\label{del_u}.
\end{eqnarray}
Substituting Eqs.~(\ref{del_m})--(\ref{del_u}) into Eq.~(\ref{num_eigen}) with
$\vec{w}=\vec{\Theta}_m$, and taking the first-order perturbation,
the error of the numerical eigenfrequency is approximated as follows:
\begin{equation}
   \delta \Omega_m \approx \frac{\delta K(\vec{\Theta}_m^*,\vec{\Theta}_m)
   -\Omega_m^2\delta M( \vec{\Theta}_m^*,\vec{\Theta}_m)}{2\Omega_m}.
   \label{error_eigenfreq}
\end{equation}
Consequently, we have $\delta \Omega_m=0$ when the numerical operators approximately satisfy
\begin{equation}
\delta K( \vec{\Theta}_m^*,\vec{\Theta}_m) \approx
   \Omega_m^2\delta M( \vec{\Theta}_m^*,\vec{\Theta}_m).
   \label{crit_gt}
\end{equation}
Dividing Eq.~(\ref{crit_gt}) by Eq.~(\ref{ext_eigen}) with $\vec{w}=\vec{\Theta}_m^*$,
Eq.~(\ref{crit_gt}) can be rewritten as follows:
\begin{equation}
   \frac{\delta M(\vec{\Theta}_m^*, \vec{\Theta}_m)}{M(\vec{\Theta}_m^*, \vec{\Theta}_m)}
   \approx
   \frac{\delta K(\vec{\Theta}_m^*, \vec{\Theta}_m)}{K(\vec{\Theta}_m^*, \vec{\Theta}_m)}.
      \label{18}
\end{equation}
In other words, the error of the numerical eigenfrequency will be minimized
when the numerical operators have modal errors of the same ratio.
\section{Numerical integration operators}\label{sec.theory_assmpt}
We consider a FEM such that
the global operators $M^{num}$ and $K^{num}$ can be written as
the superposition of local operators for respective elements, as in the SEM.
Then, $M^{num}$ and $K^{num}$ are expressed as follows:
\begin{equation}
	M^{num}(\vec{w},\vec{v})
	=  \sum_e M_e^{num}(\vec{w},\vec{v}),\hspace{1em}
   K^{num}(\vec{w}, \vec{v})=\sum_e K_e^{num}(\vec{w},\vec{v}),
\label{global_sum_local}
\end{equation}
where $M_e^{num}$ and $K_e^{num}$ are the local operators for the $e$th-element.
Further, the local operators are decomposed as follows:
\begin{eqnarray}
	M_e^{num}(\vec{w},\vec{v})
	&= & \sum_{\alpha=x,y,z} I_0^{e,num}(w_\alpha, v_\alpha) \label{kuso4.2_1}\\
   K_e^{num}(\vec{w}, \vec{v})&=&
   \sum_{\alpha,\beta=x,y,z}\left[
   I_{\alpha\beta,\lambda}^{e,num}(w_\alpha,v_\beta)
  +
  I_{\beta\alpha,\mu}^{e,num}(w_\alpha, v_\beta)
   +
   I_{\beta\beta,\mu}^{e,num}(w_\alpha, v_\alpha)\right]
   \label{kuso4.2_2}\\
	I_0^{e,num}(w, v)
	&=&\mbox{n.i.}\int_{V_e} w \rho v dV 
	\label{kuso4.2_3}
	\\
	I^{e,num}_{\alpha\beta,Z}(w, v)
	&=&\mbox{n.i.}\int_{V_e} (\partial_\alpha w) Z (\partial_\beta v) dV,
	   \label{kuso4.2_4}
\end{eqnarray}
where $V_e$ is the volume of the $e$th-element,
$w$ and $v$ are functions of $\vec{x}$,
and $Z$ is the dummy for $\lambda$ and $\mu$.

To simplify the problem,
we consider the regular Cartesian grid with elements of lengths $\Delta x =\Delta y =\Delta z=h$,
and ignore effects of element distortion.
Furthermore, we assume that
the medium is unbounded and homogeneous.
Note that this assumption
is commonly used in measurement of numerical dispersion
\cite{MB82,BS07,SO08a,SO08b,OS10}.
Indeed, in this case the errors of the numerical eigenfrequencies~(\ref{del_omega}) are exactly equivalent to the definition of numerical dispersion.
Therefore,
our objective is to derive operators which minimize the numerical dispersion.
Needless to say, from a practical point of view, we often consider computation for general heterogeneous cases,
rather than homogeneous cases.
Nevertheless, 
numerical dispersion
itself can appear as a local phenomenon in each element,
and it will be a primary factor of the numerical inaccuracy of computation even for heterogeneous cases.
This is why this analysis will effectively improve wave computation in general cases,
as shown in Section~\ref{subsec.comp_wave}.

In an unbounded homogeneous medium, 
plane waves can be used for the modes of Eq.~(\ref{ext_eigen}):
\begin{equation}
	\vec{\Theta}_m (\vec{x}) = \vec{U}_\gamma \Psi_{\vec{k}}(\vec{x}),\hspace{1em}
	\Psi_{\vec{k}}(\vec{x}) = \mathrm{e}^{\mathrm{i} \vec{k} \cdot \vec{x} },
	\label{20}
\end{equation}
where
$\vec{k} = (k_x, k_y, k_z)$ denotes the wavenumber vector,
$\gamma$ specifies the type of polarization,
$\vec{U}_\gamma=(U^\gamma_x, U^\gamma_y, U^\gamma_z)$ denotes the amplitude vector.
Note that $\vec{U}_\gamma$ is parallel to $\vec{k}$
when $\gamma$ specifies a P-wave (compressional wave)
or perpendicular to $\vec{k}$
when $\gamma$ specifies an S-wave (shear wave),
and they are orthogonal to each other.
Note also that the plane waves are characterized by $(\gamma,\vec{k})$,
and now the mode number $m$ of Eq.~(\ref{20}) stands for a pair $(\gamma,\vec{k})$

We define
the following three types of numerical integration operators:
\begin{eqnarray}
     A^{num}(\phi,\psi) &=& \mbox{n.i.} \int_{-h/2}^{h/2} \phi \psi dx
     \label{21}\\
     B^{num}(\phi,\psi) &=& \mbox{n.i.} \int_{-h/2}^{h/2} \phi' \psi' dx
     \label{22}\\
     C^{num}(\phi,\psi) &=& \mbox{n.i.} \int_{-h/2}^{h/2} \phi' \psi dx,
     \label{23}
\end{eqnarray}
where $\phi$ and $\psi$ are functions of $x$, and prime denotes the spatial differentiation.
We here suppose that
$M_e^{num}$ and $K_e^{num}$
can be expressed in terms of
tensor products of the above integration operators for the $x$, $y$, and $z$-axes.
Then, noting that the plane waves (\ref{20}) are simply expressed as
the products of harmonic functions for the $x$, $y$, and $z$-axes,
$M_e^{num}(\vec{\Theta}_m^*, \vec{\Theta}_m)$ and $K_e^{num}(\vec{\Theta}_m^*, \vec{\Theta}_m)$ can be expressed in terms of products of
quadratic forms of the integration operators
$A^{num}(p_{k_\alpha}^*, p_{k_\alpha} )$, $B^{num}(p_{k_\alpha}^*, p_{k_\alpha} )$,
and $C^{num}(p_{k_\alpha}^*, p_{k_\alpha} )$,
where $p_{k_\alpha}$ denotes the harmonic wave function which
propagates along the $\alpha$-axis:
\begin{equation}
  p_{k_\alpha} (x_\alpha)= \mathrm{e}^{\mathrm{i} k_{\alpha} x_\alpha}
   \label{25}
\end{equation}
with $k_\alpha=k_x, k_y, k_z$ and $x_\alpha=x, y, z$.
Therefore,
Eq.~(\ref{18}) can be simply decomposed into the following conditions for the integration operators:
\begin{equation}
   \frac{\delta A(p_{k_\alpha}^*, p_{k_\alpha} )}{ A(p_{k_\alpha}^*, p_{k_\alpha} ) }
   \approx
    \frac{\delta B(p_{k_\alpha}^*, p_{k_\alpha} )}{ B(p_{k_\alpha}^*, p_{k_\alpha} ) }
    \approx
    \frac{\delta C(p_{k_\alpha}^*, p_{k_\alpha} )}{ C(p_{k_\alpha}^*, p_{k_\alpha} ) }
    \approx \mathcal{E},
             \label{24}
\end{equation}
where $\mathcal{E}$ represents lower-order error terms of the integration operators,
$A$, $B$, and $C$ are the exact integration operators corresponding to
$A^{num}$, $B^{num}$, and $C^{num}$, respectively,
and $\delta A$, $\delta B$, and $\delta C$ are the errors, where
\begin{eqnarray}
     A^{num}(\phi,\psi) &=& A(\phi,\psi) +\delta A(\phi,\psi)   \label{26}\\
     B^{num}(\phi,\psi) &=& B(\phi,\psi) +\delta B(\phi,\psi)   \label{27}\\
     C^{num}(\phi,\psi) &=& C(\phi,\psi) +\delta C(\phi,\psi).   \label{28}
\end{eqnarray}
As indicated by Eq.~(\ref{18}) for this case,
the numerical dispersion will be suppressed
when the contributions of $M^{num}$ and $K^{num}$ to the dispersion are the same ratio.
Then, since they are based on the integration operators~(\ref{21})--(\ref{23}),
this situation will be realized
when the integration operators
have the contributions of the same ratio.

Now we confirm that $M^{num}$ and $K^{num}$
will approximately satisfy Eq.~(\ref{18})
when $A^{num}$, $B^{num}$, and $C^{num}$ satisfy Eq.~(\ref{24}).
We assume that $(\rho,\lambda,\mu)$ are constant
and the modes are given by plane waves~(\ref{20}).
In this case,
if the local operators for each element satisfy Eq.~(\ref{18}),
so will the global operators.
Hence, this argument can proceed with a focus on a single element $V_e=[-h/2,h/2]^3$.
Firstly, we evaluate $I_0^{e,num}(\Psi_{\vec{k}}^*, \Psi_{\vec{k}})$
based on the first-order perturbation as follows:
\begin{eqnarray}
	\nonumber
	I_0^{e,num}(\Psi_{\vec{k}}^*,\Psi_{\vec{k}})
	&=&\rho \left[ \mathrm{n.i.}
	 \int_{-h/2}^{h/2} p_{k_x}^*p_{k_x} dx\right]
	\left[ \mathrm{n.i.}
	\int_{-h/2}^{h/2} p_{k_y}^*p_{k_y} dy\right]
	\left[ \mathrm{n.i.}
	\int_{-h/2}^{h/2} p_{k_z}^*p_{k_z} dz\right]\\
	\nonumber
	&=&
	\rho A^{num}(p^*_{k_x},p_{k_x})A^{num}(p^*_{k_y},p_{k_y})A^{num}(p^*_{k_z},p_{k_z})\\
	\nonumber
	&\approx&
	\left[1+\frac{\delta A(p^*_{k_x},p_{k_x})}{A(p^*_{k_x},p_{k_x})}
	+\frac{\delta A(p^*_{k_y},p_{k_y})}{A(p^*_{k_y},p_{k_y})}
	+\frac{\delta A(p^*_{k_z},p_{k_z})}{A(p^*_{k_z},p_{k_z})}
	\right]
	\rho A(p^*_{k_x},p_{k_x})A(p^*_{k_y},p_{k_y})A(p^*_{k_z},p_{k_z})\\
	&\approx&
	(1+3\mathcal{E})I_0^e(\Psi_{\vec{k}}^*,\Psi_{\vec{k}}),
	\label{1_r1_sec2}
\end{eqnarray}
where $I_0^e$ denotes the exact operator corresponding to $I_0^{e,num}$.
Note that we omit $\delta^2$ and $\delta^3$-terms, and use the condition~(\ref{24}) to obtain the fourth line.

Next, we evaluate $I^{e,num}_{\alpha\beta,Z}(\Psi_{\vec{k}}^*, \Psi_{\vec{k}})$ for a constant $Z$.
$I^{e,num}_{\alpha\beta,Z}$ are classified into ``non-mixed-derivative'' cases $\alpha = \beta$
and ``mixed-derivative'' cases $\alpha\neq \beta$.
$I^{e,num}_{xx,Z}(\Psi_{\vec{k}}^*, \Psi_{\vec{k}})$
is evaluated based on the first-order perturbation as follows:
\begin{eqnarray}
	\nonumber
	I^{e,num}_{xx,Z}(\Psi_{\vec{k}}^*,\Psi_{\vec{k}})
	&=&
	Z\left[ \mathrm{n.i.}
	\int_{-h/2}^{h/2} (\partial_xp_{k_x})^* (\partial_x p_{k_x}) dx\right]
	\left[ \mathrm{n.i.}
	\int_{-h/2}^{h/2} p_{k_y}^*p_{k_y} dy\right]
	\left[ \mathrm{n.i.}
	\int_{-h/2}^{h/2} p_{k_z}^*p_{k_z} dz\right]\\
		\nonumber
	&=& Z B^{num}(p^*_{k_x},p_{k_x})A^{num}(p^*_{k_y},p_{k_y})A^{num}(p^*_{k_z},p_{k_z})\\
	\nonumber
	&\approx&
	\left[1+\frac{\delta B(p^*_{k_x},p_{k_x})}{B(p^*_{k_x},p_{k_x})}
	+\frac{\delta A(p^*_{k_y},p_{k_y})}{A(p^*_{k_y},p_{k_y})}
	+\frac{\delta A(p^*_{k_z},p_{k_z})}{A(p^*_{k_z},p_{k_z})}
	\right]
	 Z B(p^*_{k_x},p_{k_x})A(p^*_{k_y},p_{k_y})A(p^*_{k_z},p_{k_z})\\
	&\approx&
	(1+3\mathcal{E})I_{xx,Z}^e(\Psi_{\vec{k}}^*,\Psi_{\vec{k}}),
	\label{4_r1_sec2}
\end{eqnarray}
where $I_{\alpha\beta,Z}^e$ denotes the exact operator corresponding to $I_{\alpha\beta,Z}^{e,num}$.
Similarly,
$I^{e,num}_{xy,Z}(\Psi_{\vec{k}}^*, \Psi_{\vec{k}})$
is evaluated as follows:
\begin{eqnarray}
	\nonumber
	I^{e,num}_{xy,Z}(\Psi_{\vec{k}}^*,\Psi_{\vec{k}})
	&=&Z\left[ \mathrm{n.i.}
	\int_{-h/2}^{h/2} (\partial_xp_{k_x})^* p_{k_x} dx\right]
	\left[ \mathrm{n.i.}
	\int_{-h/2}^{h/2} p_{k_y}^* (\partial_y p_{k_y}) dy\right]
	\left[\mathrm{n.i.}
	\int_{-h/2}^{h/2} p_{k_z}^*p_{k_z} dz\right]\\
		\nonumber
	&=& ZC^{num}(p^*_{k_x},p_{k_x})C^{num}(p_{k_y},p_{k_y}^*)A^{num}(p^*_{k_z},p_{k_z})\\
	\nonumber
	&\approx&
	\left[1+\frac{\delta C(p^*_{k_x},p_{k_x})}{C(p^*_{k_x},p_{k_x})}
	+\frac{\delta C(p_{k_y},p^*_{k_y})}{C(p_{k_y},p^*_{k_y})}
	+\frac{\delta A(p^*_{k_z},p_{k_z})}{A(p^*_{k_z},p_{k_z})}
	\right]
	 ZC(p^*_{k_x},p_{k_x})C(p_{k_y},p^*_{k_y})A(p^*_{k_z},p_{k_z})\\
	&\approx&
	(1+3\mathcal{E})I^e_{xy,Z}(\Psi_{\vec{k}}^*, \Psi_{\vec{k}}).
	\label{5_r1_sec2}
\end{eqnarray}
We will also have the same result for the other cases of $(\alpha,\beta)$,
and thus we see that
\begin{equation}
	I^{e,num}_{\alpha\beta,Z}(\Psi_{\vec{k}}^*,\Psi_{\vec{k}})
	\approx (1+3\mathcal{E})I^e_{\alpha\beta,Z}(\Psi_{\vec{k}}^*,\Psi_{\vec{k}}).
	\label{6_r1_sec2}
\end{equation}
Consequently, substituting $(\vec{w},\vec{v})=(\vec{\Theta}_m^*,\vec{\Theta}_m)$
into Eqs.~(\ref{kuso4.2_1}) and (\ref{kuso4.2_2}) and comparing them,
we have
\begin{equation}
	\frac{\delta M_e(\vec{\Theta}_m^*,\vec{\Theta}_m)}{M_e(\vec{\Theta}_m^*,\vec{\Theta}_m)}
	\approx
	\frac{\delta K_e(\vec{\Theta}_m^*,\vec{\Theta}_m)}{K_e(\vec{\Theta}_m^*,\vec{\Theta}_m)}
	\approx
	3\mathcal{E},
	\label{7_r1_sec2}
\end{equation}
and thus we approximately obtain Eq.~(\ref{18}).

In the following subsections, we introduce the SEM integration operators, and then modified integration operators
which satisfy Eq.~(\ref{24}). 
Let us consider a harmonic wave function $p_{k_x}(x)$ in the domain of integration $[-h/2,h/2]$
along the $x$-axis.
We note that there is no other measure to define the scale of space than the length  of the domain $h$
and the wavelength of the harmonic wave function $\ell=2\pi/k_x$;
i.e., to increase the wavelength is to decrease the domain size.
Therefore, hereafter in this section,
we fix the value of $h$ as $h=2$ without loss of generality,
and we define the following scale factor instead of wavenumber $k_x$:
\begin{equation}
\hat{h}=hk_x/2=h \pi /\ell.
          \label{29}
\end{equation}
If necessary, the reader can proceed following discussions
for general cases of arbitrary values of $h$,
by multiplying the integrals (\ref{21})--(\ref{23}) with $h=2$
by factors of $h/2$, $2/h$, and 1, respectively,
and by replacing $\phi(x)$ and $\psi(x)$ with $\phi(hx/2)$
and $\psi(hx/2)$.

\subsection{SEM integration operators}\label{subsec.theory_ordinal}
We denote numerical operators $A^{num}$, $B^{num}$, and $C^{num}$
defined based on the SEM by $A^{SEM}$, $B^{SEM}$, and $C^{SEM}$,
respectively.
For the SEM,
the operands of Eqs.~(\ref{21})--(\ref{23}) are approximated as follows:
\begin{eqnarray}
   \phi(x)&\approx& \sum_{i=0}^n \phi(x_i) L_i(x)   
    \label{30}\\
   \psi(x) &\approx& \sum_{i=0}^n \psi(x_i) L_i(x),
       \label{31}
\end{eqnarray}
where $n$ is the degree of the polynomial (i.e. the order of an element),
$x_i$ are
the GLL nodes in ascending order as the index $i$ increases,
and
$L_i$ are Lagrange interpolating polynomials defined as
\begin{equation}
   L_i(x)=\prod_{j=0, j\neq i}^n \frac{x-x_j}{x_i-x_j}.
       \label{32}
\end{equation}
Then, the SEM integration operators are defined based on the GLL quadrature rule as follows:
\begin{eqnarray}
\nonumber
   A^{SEM}(\phi,\psi) &=& \sum_{i,j=0}^n \phi(x_i) \psi(x_j) \int_{-1}^1 L_i L_j dx\\
   &\approx& \sum_{i=0}^n \phi_i q_i \psi_i
          \label{33}\\
\nonumber
   B^{SEM}(\phi,\psi) &=& \sum_{i,j=0}^n \phi(x_i) \psi(x_j) \int_{-1}^1 L_i' L_j' dx\\
   &=& \sum_{r=0}^n  \left[ \sum_{i=0}^nD_{ri}\phi_i \right] q_r \left[ \sum_{j=0}^n D_{rj}\psi_j \right]
          \label{34}\\
\nonumber
   C^{SEM}(\phi,\psi) &=& \sum_{i,j=0}^n \phi(x_i) \psi(x_j) \int_{-1}^1 L_i' L_j dx\\
   &=& \sum_{i,j=0}^n D_{ji} \phi_i q_j  \psi_j,
          \label{35}
\end{eqnarray}
where $\phi_i=\phi(x_i)$, $\psi_i=\psi(x_i)$, $q_i$ are the GLL weights,
and
\begin{equation}
   D_{ij}=L_j'(x_i).
   \label{36}
\end{equation}
Explicit expressions of $q_i$ and $L_j'(x_i)$ are given by Eqs.~(\ref{a8}) and ({\ref{a21}) in \ref{appendix_a}.
Note that the integral in Eq.~(\ref{33}) is approximated based on the GLL rule, 
whereas those in Eqs.~(\ref{34}) and (\ref{35}) are exactly calculated,
since the GLL rule exactly computes an integral when the integrand is a polynomial of degree $(2n-1)$ or below.

For the exact operators, we have
\begin{eqnarray}
   A(p_{\hat{h}}^*,p_{\hat{h}}) &=&\int_{-1}^1 \left|p_{\hat{h}}(x)\right|^2dx =2
   \label{37}\\
   B(p_{\hat{h}}^*,p_{\hat{h}}) &=&\int_{-1}^1 \left|p_{\hat{h}}'(x)\right|^2dx =2\hat{h}^2
   \label{38}\\
   C(p_{\hat{h}}^*,p_{\hat{h}}) &=&\int_{-1}^1\left[p_{\hat{h}}'(x)\right]^* p_{\hat{h}}(x) dx =-2\mathrm{i} \hat{h}.
   \label{39}
\end{eqnarray}
Substituting $(p_{\hat{h}}^*,p_{\hat{h}})$ into $(\phi, \psi)$ of Eqs.~(\ref{33})--(\ref{35}), respectively,
and comparing them with the exact results (\ref{37})--(\ref{39}),
the relative errors of the SEM integration operators are given by
\begin{eqnarray}
   \frac{\delta A^{SEM}(p_{\hat{h}}^*,p_{\hat{h}})}
   {A(p_{\hat{h}}^*,p_{\hat{h}})}&=&0
   \label{40}\\
   \frac{\delta B^{SEM}(p_{\hat{h}}^*,p_{\hat{h}})}
   {B(p_{\hat{h}}^*,p_{\hat{h}})} &=&
   \mathcal{F}_n \hat{h}^{2n} +O(\hat{h}^{2n+2})
   \label{41}\\
   \frac{\delta C^{SEM}(p_{\hat{h}}^*,p_{\hat{h}})}
   {C(p_{\hat{h}}^*,p_{\hat{h}})} &=&
        (n+1)\mathcal{F}_n \hat{h}^{2n} +O(\hat{h}^{2n+1}),
   \label{42}
\end{eqnarray}
where $O(\hat{h}^{l})$ represents terms having $\hat{h}$ to the power of $l$ or above,
and $\mathcal{F}_n$ is given by
\begin{equation}
   \mathcal{F}_n = -\frac{n}{4(2n+1)(n!)^2} \left[\sum_{i=0}^nq_iP_n(x_i)x_i^n \right]^2,
      \label{43}
\end{equation}
where $P_n$ is the Legendre polynomial of the $n$th-order.
We show their derivations in \ref{appendix_b}.
Comparing Eqs.~(\ref{40})--(\ref{42}),
we obviously see that
their $2n$th-order terms
are different from each other,
and thus
the SEM operators (\ref{33})--(\ref{35})
do not satisfy Eq.~(\ref{24}) at the $2n$th-order.
\subsection{Modified integration operators}\label{subsec.theory_modify}
Hereafter, we denote numerical operators $A^{num}$, $B^{num}$, and $C^{num}$ such that 
they satisfy Eq.~(\ref{24}) by $A^{opt}$, $B^{opt}$, and $C^{opt}$, respectively.
Firstly, we assume
that $B^{opt}$ is simply given by
\begin{equation}
   B^{opt}(\phi,\psi)=B^{SEM}(\phi,\psi),
      \label{44}
\end{equation}
and then we define $A^{opt}$ and $C^{opt}$
such that their errors have the same ratio as Eq.~(\ref{41}).
Note that in this paper we propose one solution,
while there could be other ways of modification
which satisfy Eq.~(\ref{24}), but do not assume Eq.~(\ref{44}).
The starting point is exactly same as the mass-blending approaches~(e.g. \cite{M84,SO07,AW10}).

In order to define $A^{opt}$,
we make another FEM definition for $A^{num}$
such that we compute the integral in Eq.~(\ref{33}) exactly as follows:
\begin{eqnarray}
\nonumber
   A^{FEM}(\phi,\psi) &=& \sum_{i,j=0}^n \phi_i \psi_j \int_{-1}^1 L_i L_j dx\\
   &=& A^{SEM}(\phi,\psi) -\frac{n(n+1)}{2(2n+1)}
 \left[\sum_{i=0}^n \phi_i q_i P_n(x_i) \right]\left[\sum_{j=0}^n \psi_j q_jP_n(x_j)\right],
          \label{45}
\end{eqnarray}
where $A^{SEM}$ represents the second line of Eq.~(\ref{33}).
An integration formula for the derivation of the second line
is shown in Eq.~(\ref{a23}) in \ref{appendix_a}.
The relative error of $A^{FEM}(p_{\hat{h}}^*,p_{\hat{h}})$ is given by
\begin{equation}
   \frac{\delta A^{FEM}(p_{\hat{h}}^*,p_{\hat{h}})}
   {A(p_{\hat{h}}^*,p_{\hat{h}})}=
     (n+1)\mathcal{F}_n \hat{h}^{2n} +O(\hat{h}^{2n+2}).
   \label{46}
\end{equation}
The derivation is shown in \ref{appendix_b}
(see eq.~\ref{b2}).

We define $A^{opt}$ by blending $A^{SEM}$ and $A^{FEM}$ of
Eqs.~(\ref{33}) and (\ref{45}).
Since
the error of $A^{opt}$ is also linearly controlled by its blending ratio,
we immediately obtain the optimal ratio such that the error is equal to Eq.~(\ref{41})
as follows:
\begin{eqnarray}
   \nonumber
  A^{opt}(\phi,\psi) &=& \frac{n}{n+1}A^{SEM}(\phi,\psi)
  +  \frac{1}{n+1} A^{FEM}(\phi,\psi)\\
   &=& A^{SEM}(\phi,\psi) -\frac{n}{2(2n+1)}
   \left[\sum_{i=0}^n \phi_i q_i P_n(x_i) \right]
   \left[\sum_{j=0}^n \psi_j q_jP_n(x_j)\right].
   \label{47}
\end{eqnarray}
Note that the blending approach itself
has already been reported by previous papers,
and the optimal value of blending ratio has been analytically given by \cite{AW10}.
Eq.~(\ref{47}) follows their result, except for differences in the derivation.

To define $C^{opt}$,
we make the following approximation for the operand $\phi$, instead of Eq.~(\ref{30}):
\begin{equation}
   \phi(x) \approx \sum_{i=-1}^n \phi(x_i) X_i(x),
      \label{48}
\end{equation}
where $X_i$ are Lagrange interpolating polynomials of degree $(n+1)$ defined as follows:
\begin{equation}
X_i(x) = \prod_{j=-1, j\neq i}^n \frac{x-x_j}{x_i-x_j}.
      \label{49}
\end{equation}
Note that the product is taken for $j$ from $-1$ to $n$ except $j = i$.
$x_i$ denote
the GLL nodes except for the case of $i = -1$:
$x_{-1}$ is an arbitrary node other than $x_0,x_1,\dots, x_n$.
In order to keep the total number of grid points,
we use a node already existing outside the domain of the integration $[-1,1]$
as $x_{-1}$.
In this paper, we define $x_{-1}$ as follows:
\begin{equation}
   x_{-1} = x_{n-1} - 2.
         \label{50}
\end{equation}
The definition (\ref{50}) indicates
that $x_{-1}$ is located next to the left of $x_0$,
when we take the positive direction to the right of the $x$-axis.
In other words, the node (\ref{50}) is the $(n-1)$th GLL node defined on $[-3,-1]$,
and then it is shared by the two adjacent elements to numerically compute the integral (\ref{23}) for each domain.
Note also that, by the definition, the approximation (\ref{48})
cannot be used 
when the domain is located on the leftmost of a medium.
We can use a node on the right side for that case,
or may simply
use the ordinary SEM operator $C^{SEM}$ only for that case.
In contrast to $\phi$, we use the same approximation as Eq.~(\ref{31}) for the operand $\psi$.
Then, we define $C^{opt}$ as follows:
\begin{eqnarray}
\nonumber
   C^{opt}(\phi,\psi) &=&
   \sum_{i=-1, j=0}^n \phi_i \psi_j \int_{-1}^1 X'_i L_j dx\\
   \nonumber
   &=& C^{SEM}(\phi,\psi)+
   \frac{n^2(n+1)}{2(2n+1)}
    \left[ 
   \sum_{i=0}^n\frac{\phi_iq_i P_n(x_i)}{ x_i-x_{-1}}
   + \frac{2\phi_{-1}}{(x_{-1}^2-1)P'_n(x_{-1})} \right]
    \left[\sum_{j=0}^n\psi_jq_j P_n(x_j)\right],\\
            \label{51}
\end{eqnarray}
where $\phi_{-1}=\phi(x_{-1})$.
Integration formulas for the derivation of the second line
are shown in Eqs.~(\ref{a25}) and (\ref{a26}) in \ref{appendix_a}.

The relative errors of the modified integration operators are given by
\begin{eqnarray}
   \frac{\delta A^{opt}(p_{\hat{h}}^*, p_{\hat{h}})}{A(p_{\hat{h}}^*, p_{\hat{h}})}
   &=&\mathcal{F}_n \hat{h}^{2n} + O(\hat{h}^{2n+2})
   \label{52}\\
   \frac{\delta B^{opt}(p_{\hat{h}}^*, p_{\hat{h}})}{B(p_{\hat{h}}^*, p_{\hat{h}})}
   &=&\mathcal{F}_n \hat{h}^{2n} + O(\hat{h}^{2n+2})
   \label{53}\\
   \frac{\delta C^{opt}(p_{\hat{h}}^*, p_{\hat{h}})}{C(p_{\hat{h}}^*, p_{\hat{h}})}
   &=&\mathcal{F}_n \hat{h}^{2n} + O(\hat{h}^{2n+1}).
   \label{54}
\end{eqnarray}
Their derivations are shown in \ref{appendix_b}.
Comparing Eqs.~(\ref{52})--(\ref{54}),
we obviously see that
the modified operators
satisfy Eq.~(\ref{24}) at the $2n$th-order.
However, in contrast to the cases of $A^{opt}$ and $B^{opt}$,
the relative error of $C^{opt}(p_{\hat{h}}^*, p_{\hat{h}})$
may have a non-zero pure imaginary term at the $(2n+1)$th-order.
Nevertheless,
$\delta K_e(\vec{\Theta}_m^*,\vec{\Theta}_m )$ will not be affected by
the $(2n+1)$th-order imaginary error
regardless of its value,
since it is canceled out with its complex conjugate when we construct
$K_e^{num}(\vec{\Theta}_m^*,\vec{\Theta}_m)$.
Consequently, the modified operators will satisfy Eq.~(\ref{18})
up to the $(2n+1)$th-order.
\section{Construction of mass and stiffness matrices}\label{sec.3-Dops}
\subsection{Matrix form of modified operators}\label{sec4.1}
We express
$A^{opt}$, $B^{opt}$, and
$C^{opt}$ in terms of matrix products as follows:
\begin{eqnarray}
   A^{opt}(\phi, \psi)
   &=& \sum_{i,j=0}^n \phi_i A_{ij}^{opt} \psi_j
         \label{3.1}\\
   B^{opt}(\phi, \psi)
   &=& \sum_{i,j=0}^n \phi_i B_{ij}^{opt} \psi_j
         \label{3.2}\\
   C^{opt}(\phi, \psi)
   &=& \sum_{i=-1, j=0}^n \phi_i C_{ij}^{opt} \psi_j.
         \label{3.3}
\end{eqnarray}
Note that the size of matrices $\left(A_{ij}^{opt} \right)$ and $\left(B_{ij}^{opt} \right)$
is $(n+1)\times (n+1)$, whereas the size of $\left(C_{ij}^{opt} \right)$ is $(n+2)\times (n+1)$.
The components of these matrices are
\begin{eqnarray}
  A^{opt}_{ij} &=&
   A^{SEM}_{ij}
   -\frac{n}{2(2n+1)} b_ib_j
          \label{3.5}\\
     B^{opt}_{ij} &=& B^{SEM}_{ij}
          \label{3.6}\\
     C^{opt}_{ij} &=&
   C^{SEM}_{ij}
   +\frac{n^2(n+1)}{2n+1} s_i b_j,
          \label{3.7}
\end{eqnarray}
where $i,j=0,\dots,n$, except that Eq.~(\ref{3.7}) includes the cases of $i=-1$ with $C^{SEM}_{-1 j}=0$,
$b_i$ and $s_i$ are
\begin{eqnarray}
 b_i &=& 
   q_i P_n(x_i)
   \label{re_b_i}\\
 s_i &=&   \left\{
   \begin{array}{cl}
   \displaystyle{
   \frac{1}{2}
   \frac{q_i P_n(x_i)}{x_i-x_{-1}} }
   & \mbox{if}\hspace{0.5em} i=0,\dots,n\\
   \vspace{-0.8em}\\
            \displaystyle{
   \frac{1}{(x_{-1}^2-1)P'_n(x_{-1})}} &
   \mbox{if}\hspace{0.5em}  i = -1,
   \end{array}
   \right.
   \label{re_s_i}
\end{eqnarray}
and $A_{ij}^{SEM}$, $B_{ij}^{SEM}$, 
and $C_{ij}^{SEM}$ ($i>-1$) are the components of the $(n+1)\times (n+1)$ matrix operators of the SEM:
\begin{eqnarray}
  A^{SEM}_{ij} &=&
   q_i \delta_{ij}
          \label{3.5_SEM}\\
     B^{SEM}_{ij} &=&
     \sum_{r=0}^n D_{ri} q_r D_{rj}
          \label{3.6_SEM}\\
     C^{SEM}_{ij} &=&  D_{ji}q_j.
          \label{3.7_SEM}
\end{eqnarray}

\subsection{Homogeneous case}\label{sec4.2}
We firstly define numerical operators $I^{e,num}_0$ and $I_{\alpha\beta,Z}^{e,num}$ of Eqs.~(\ref{kuso4.2_3}) and (\ref{kuso4.2_4})
for constant values of $(\rho,\lambda,\mu)$
by using the matrix operators defined in Section~\ref{sec4.1},
and thereby derive the explicit forms of $M_e^{num}$ and $K_e^{num}$ of Eqs.~(\ref{kuso4.2_1}) and (\ref{kuso4.2_2}) for a homogeneous case.
For simplicity, here we consider the 2-D case,
but it can be straightforwardly extended to the 3-D case as the following derivation.
Now we focus on element $V_e=[-h/2,h/2]^2$.
The global operators are obtained by assembling
local operators defined as the following description for each element,
except that we use the ordinary SEM operators only for the boundary elements,
since $x_{-1}$ of Eq.~(\ref{50}) cannot be defined for such elements.

Operator $I_0^{e,num}$ is defined by applying Eq.~(\ref{3.1}) one by one
for the $x$ and $y$-axes as follows:
\begin{eqnarray}
\nonumber
   I_0^{e,num}(w, v) &=&
   \rho\,\mbox{n.i.} \int_{-h/2}^{h/2} \left[\mbox{n.i.} \int_{-h/2}^{h/2}
   w(x,y) v(x,y) dx \right]dy\\
   &=&
   \rho\left(\frac{h}{2}\right)^2 
   \sum_{i_x,j_x=0}^n  \sum_{i_y,j_y=0}^n
   w_{(i_x,i_y)} v_{(j_x,j_y)}
   A^{opt}_{i_xj_x} A^{opt}_{i_yj_y},
                 \label{3.15}
\end{eqnarray}
where
$w_{(i_x,i_y)}=w(hx_{i_x}/2,hx_{i_y}/2)$ and
$v_{(i_x,i_y)}=v(hx_{i_x}/2,hx_{i_y}/2)$.
Similarly, $I_{xx,Z}^{e,num}$ is defined by
\begin{eqnarray}
\nonumber
   I_{xx,Z}^{e,num}(w,v) &=&
   Z\, \mbox{n.i.} \int_{-h/2}^{h/2} \left[
   \mbox{n.i.} \int_{-h/2}^{h/2}
   \partial_x w(x,y)\partial_x v(x,y) dx \right] dy\\
   &=&
   Z\sum_{i_x,j_x=0}^n  \sum_{i_y,j_y=0}^n
   w_{(i_x,i_y)} v_{(j_x,j_y)}
   B^{opt}_{i_xj_x} A^{opt}_{i_yj_y},
       \label{3.16}
\end{eqnarray}
and $I_{xy,Z}^{e,num}$ is
\begin{eqnarray}
\nonumber
   I_{xy,Z}^{e,num}(w, v) &=&
      Z\, \mbox{n.i.} \int_{-h/2}^{h/2} \left[
      \mbox{n.i.} \int_{-h/2}^{h/2} \partial_x w(x,y) \partial_y v(x,y) dx \right]dy\\
   &=&
     Z \sum_{i_x=-1, j_x=0}^n  \sum_{i_y=0, j_y=-1}^n
   w_{(i_x,i_y)}  v_{(j_x,j_y)}
   C^{opt}_{i_xj_x} C^{opt}_{j_yi_y}.
       \label{3.19}
\end{eqnarray}
$I_{yy,Z}^{e,num}$ and $I_{yx,Z}^{e,num}$ are also defined in similar ways.

Finally, $M_e^{num}$ and $K_e^{num}$ are defined as follows:
\begin{equation}
	M^{num}_e(\vec{w},\vec{v})
	=\left(\begin{array}{c} \mathbf{w}_x^e \\ \mathbf{w}_y^e \end{array}\right)^T
	\mathbf{M}_e
	\left(\begin{array}{c} \mathbf{v}_x^e \\\mathbf{v}_y^e \end{array}\right),\hspace{1em}
	K^{num}_e(\vec{w},\vec{v})
	=\left(\begin{array}{c} \mathbf{w}_x^e \\ \mathbf{w}_y^e \end{array}\right)^T
	\mathbf{K}_e
	\left(\begin{array}{c} \mathbf{v}_x^e \\\mathbf{v}_y^e \end{array}\right),
	\label{kuso4.2_5}
\end{equation}
where $\mathbf{w}^e_\alpha$ and $\mathbf{v}^e_\alpha$ denote $(n+2)^2$-dimensional vectors
whose components are
$ \left(\mathbf{w}^e_\alpha \right)_{(i_x,i_y)} =w_\alpha(hx_{i_x}/2,hx_{i_y}/2)$
and $\left( \mathbf{v}^e_\alpha\right)_{(i_x,i_y)}=v_\alpha(hx_{i_x}/2,hx_{i_y}/2)$,
respectively, and
$\mathbf{M}_e$ and $\mathbf{K}_e$ are the local mass and stiffness matrices
given by
\begin{eqnarray}
  \mathbf{M}_e &= &\left(
	\begin{array}{cc}
	\mathbf{I}^e_0 & \mathbf{0} \\
	 \mathbf{0} & \mathbf{I}^e_0
	\end{array}
	\right)
	\label{kuso4.2_7}\\
	\mathbf{K}_e
	&=&
	\left(
	\begin{array}{cc}
	\mathbf{I}^e_{xx,(\lambda+2\mu)}+\mathbf{I}^e_{yy,\mu} &
	\mathbf{I}^e_{xy,\lambda} + \mathbf{I}^e_{yx,\mu}\\
	\mathbf{I}^e_{yx,\lambda} + \mathbf{I}^e_{xy,\mu}&
	\mathbf{I}^e_{yy,(\lambda+2\mu)} + \mathbf{I}^e_{xx,\mu}
	\end{array}
	\right)
	\label{kuso4.2_8}
\end{eqnarray}
and $\mathbf{I}^e_0$ and $\mathbf{I}^e_{\alpha\beta,Z}$ are matrices of sizes $(n+2)^2\times (n+2)^2$
whose components are 
\begin{eqnarray}
	\left(\mathbf{I}^e_0\right)_{(i_x,i_y)(j_x,j_y)} &=&  (h^2/4)\rho A_{i_x j_x}^{opt} A_{i_y j_y}^{opt}
	\label{kuso4.2_9} \\
	\left( \mathbf{I}^e_{xx,Z} \right)_{(i_x,i_y)(j_x,j_y)}
	&=&  Z B_{i_x j_x}^{opt} A_{i_y j_y}^{opt}
	\label{kuso4.2_10} \\
	\left( \mathbf{I}^e_{yy,Z} \right)_{(i_x,i_y)(j_x,j_y)}
	&=&  Z A_{i_x j_x}^{opt} B_{i_y j_y}^{opt}
	\label{kuso4.2_11} \\
	\left(\mathbf{I}^e_{xy,Z}\right)_{(i_x,i_y)(j_x,j_y)}
	&=& Z C_{i_x j_x}^{opt} C_{j_y i_y}^{opt}
	\label{kuso4.2_12} \\
	\left(\mathbf{I}^e_{yx,Z}\right)_{(i_x,i_y)(j_x,j_y)}
	 &=& Z C_{j_x i_x}^{opt} C_{i_y j_y}^{opt},
	\label{kuso4.2_13}
\end{eqnarray}
where $i_x,i_y,j_x,j_y=0,\dots,n$
except that $i_x$ and $j_y$ of Eq.~(\ref{kuso4.2_12}) and $j_x$ and $i_y$ of Eq.~(\ref{kuso4.2_13})
are taken from $-1$ to $n$.
Note that, for convenience of notation,
we define the sizes of these matrices uniformly as $(n+2)^2\times (n+2)^2$,
with components undefined in Eqs.~(\ref{kuso4.2_9})--(\ref{kuso4.2_13}) set as zero.

Note that we implicitly use different types of shape functions
as (\ref{32}) and (\ref{49}) for numerical integrations (\ref{3.15})--(\ref{3.19}).
Nevertheless,
the operands of global operators
$M^{num}$ and $K^{num}$
can be represented
by a unique set of global nodal basis functions,
each of which has the value 1 at a GLL node in an element
and zero at all other nodes.
This is because both shape functions (\ref{32}) and (\ref{49})
are also nodal functions based on the GLL nodes,
and thus numerical integrations (\ref{3.15})--(\ref{3.19})
need only the values of their operands at the GLL nodes to compute them.
Therefore, substituting the global basis functions into the operands
of $M^{num}$ and $K^{num}$,
we will obtain the global mass and stiffness matrices,
respectively,
which are exactly equal to those
obtained by assembling $\mathbf{M}_e$ and $\mathbf{K}_e$ of Eqs.~(\ref{kuso4.2_7}) and (\ref{kuso4.2_8})
defined for each element.
\subsection{Heterogeneous case}
We consider a case where $(\rho,\lambda,\mu)$ are continuous functions of $(x,y)$.
$I_0^{e,num}$ and $I^{e,num}_{\alpha\beta,Z}$ for the mixed derivative cases
are straightforwardly extended for this case as follows:
\begin{eqnarray}
\nonumber
   I_0^{e,num}(w, v) &=&
   \mbox{n.i.} \int_{-h/2}^{h/2} \left[ \mbox{n.i.} \int_{-h/2}^{h/2}
   w(x,y)\left\{ \rho(x,y) v(x,y) \right\} dx\right] dy\\
    &=&
   \left(\frac{h}{2}\right)^2 
   \sum_{i_x,j_x=0}^n  \sum_{i_y,j_y=0}^n
   w_{(i_x,i_y)} v_{(j_x,j_y)} \rho_{(j_x,j_y)}
   A^{opt}_{i_xj_x} A^{opt}_{i_yj_y}
     \label{re_4.3_1}
\end{eqnarray}
\begin{eqnarray}
\nonumber
   I_{xy,Z}^{e,num}(w,v) &=&
      \mbox{n.i.} \int_{-h/2}^{h/2} \left[ \mbox{n.i.} \int_{-h/2}^{h/2} 
   \partial_x w(x,y) \left\{ Z(x,y)\partial_y v(x,y)\right\} dx \right]dy\\
   &=&
      \sum_{i_x=-1, j_x=0}^n  \sum_{i_y=0, j_y=-1}^n
   w_{(i_x,i_y)}  v_{(j_x,j_y)} Z_{(j_x,i_y)}
   C^{opt}_{i_xj_x} C^{opt}_{j_yi_y},
     \label{re_4.3_2}
\end{eqnarray}
where $\rho_{(i_x,i_y)} = \rho(hx_{i_x}/2,hx_{i_y}/2)$
and $Z_{(i_x,i_y)}=Z(hx_{i_x}/2,hx_{i_y}/2)$.
$I_{yx,Z}^{e,num}$ is defined similarly as in Eq.~(\ref{re_4.3_2}).

In order to define $I_{\alpha\beta,Z}^{e,num}$ when $\alpha=\beta$, 
we use the following numerical integration rule, instead of $B^{opt}\,(=B^{SEM})$:
\begin{equation}
	\mbox{n.i.}\int_{-1}^1 \phi' \zeta \psi' dx =
	     \sum_{r=0}^n \left[\sum_{i=0}^n D_{ri} \phi_i\right]
	     q_r\zeta_r \left[\sum_{j=0}^nD_{rj}\psi_j\right]
	          \label{re_4.3_3}
\end{equation}
with a function $\zeta(x)$ and $\zeta_i=\zeta(x_i)$.
In this numerical integration, we use the Lagrange interpolations with the GLL nodes for $\phi$ and $\psi$
as shown in Eqs.~(\ref{30}) and (\ref{31}), and then apply the GLL quadrature rule, which leads the accuracy
of the integration~(\ref{re_4.3_3}) up to the $(2n-1)$th-order of polynomial degree
of the integrand.
Note that the definition is exactly equal to $B^{SEM}$ of Eq.~(\ref{34}) with a constant $\zeta$ multiplied
when $\zeta$ is independent to $x$.
Using this rule, $I_{xx,Z}^{e,num}$ is defined as follows:
\begin{eqnarray}
\nonumber
   I_{xx,Z}^{e,num}(w, v) &=&
      \mbox{n.i.} \int_{-h/2}^{h/2} \left[ \mbox{n.i.} \int_{-h/2}^{h/2} 
   \partial_x w(x,y) Z(x,y)\partial_x v(x,y) dx \right]dy\\
   &=&
   \sum_{i_x, j_x=0}^n  \sum_{i_y, j_y=0}^n \sum_{r_x=0}^n 
   w_{(i_x,i_y)}  Z_{(r_x,j_y)}
	    v_{(j_x,j_y)} D_{r_x i_x}  q_{r_x} D_{r_xj_x} A^{opt}_{i_yj_y}.
     \label{re_4.3_4}
\end{eqnarray}
$I_{yy,Z}^{e,num}$ is similarly defined by using Eq.~(\ref{re_4.3_3}).

Finally, the local mass and stiffness matrices $\mathbf{M}_e$ and $\mathbf{K}_e$ for the heterogeneous case are defined by replacing the submatrices (\ref{kuso4.2_9})--(\ref{kuso4.2_13}) by
\begin{eqnarray}
	\left( \mathbf{I}_0^e\right)_{(i_x,i_y)(j_x,j_y)} &=& 
	(h^2/4)\rho_{(j_x,j_y)} A_{i_x j_x}^{opt} A_{i_y j_y}^{opt}
	\label{kuso4.3_9} \\
	\left( \mathbf{I}^e_{xx,Z} \right)_{(i_x,i_y)(j_x,j_y)}
	&=&  
	\sum_{r_x=0}^n Z_{(r_x,j_y)} D_{r_x i_x}  q_{r_x} D_{r_xj_x} A^{opt}_{i_yj_y}
	\label{kuso4.3_10} \\
	\left( \mathbf{I}^e_{yy,Z} \right)_{(i_x,i_y)(j_x,j_y)}
	&=& A^{opt}_{i_xj_x}
	\sum_{r_y=0}^n Z_{(j_x, r_y)} D_{r_y i_y}  q_{r_y} D_{r_yj_y}
	\label{kuso4.3_11} \\
	\left( \mathbf{I}^e_{xy,Z}\right)_{(i_x,i_y)(j_x,j_y)}
	&=& Z_{(j_x,i_y)} C_{i_x j_x}^{opt} C_{j_y i_y}^{opt}
	\label{kuso4.3_12} \\
	\left( \mathbf{I}^e_{yx,Z}\right)_{(i_x,i_y)(j_x,j_y)}
	 &=& Z_{(i_x,j_y)} C_{j_x i_x}^{opt} C_{i_y j_y}^{opt}.
	\label{kuso4.3_13}
\end{eqnarray}
Note that the definitions of matrices~(\ref{kuso4.3_9})--(\ref{kuso4.3_13}) are consistent with those for the homogeneous case, since they are equal to Eqs.~(\ref{kuso4.2_9})--(\ref{kuso4.2_13}) when $\rho$ and $Z$ are constants. Note also that the mass and stiffness matrices of the SEM
are defined by replacing $A^{opt}_{ij}$ and $C^{opt}_{ij}$ in Eqs.~(\ref{kuso4.3_9})--(\ref{kuso4.3_13})
by $A^{SEM}_{ij}$ and $C^{SEM}_{ij}$ for the SEM, where the $i=-1$ cases are not considered.
\section{Numerical examples}\label{sec.example}
\subsection{Dispersion analysis}\label{subsec.numer_disp}
We conduct numerical dispersion analysis based on
the Rayleigh quotient approach developed by \cite{SO07,SO08a,SO08b,OS10},
which enables efficient estimation of the numerical dispersion
in a practical range of the number of grid points.
In this analysis,
we assume a time-harmonic plane wave ansatz
$(u^{num}_x,u^{num}_y)
  =(U_x,U_y)\Psi_{\vec{k}}\mathrm{e}^{- \mathrm{i}\omega_{\vec{k}} t}$
with an unknown vector $(U_x, U_y)$,
as a solution of the semi-discrete wave equation~(\ref{1})
for a 2-D unbounded homogeneous medium with no external force.
Then, considering two cases where $(w_x,w_y)=(\Psi_{\vec{k}},0)$ and
$(w_x,w_y)=(0, \Psi_{\vec{k}})$,
we obtain the following $2\times 2$ eigenvalue problem for a single element
$V_e=[-h/2,h/2]^2$:
\begin{equation}
	\left(
	\begin{array}{cc}
	\tilde{I}_{xx,(\lambda+2\mu)}
	+
	\tilde{I}_{yy,\mu} &
	\tilde{I}_{xy, \lambda}
	+
	\tilde{I}_{yx, \mu} \\
	\tilde{I}_{yx, \lambda}
	+
	\tilde{I}_{xy, \mu} &
	\tilde{I}_{yy,(\lambda+2\mu)}
	+
	\tilde{I}_{xx,\mu}
	\end{array}
	\right)
	\left(
		\begin{array}{c}
		U_x \\U_y
		\end{array}
	\right)
	=\omega_{\vec{k}}^2
		\left(
	\begin{array}{cc}
	\tilde{I}_0 & 0\\
	0 & \tilde{I}_0
	\end{array}
	\right)
		\left(
		\begin{array}{c}
		U_x \\U_y
		\end{array}
	\right),
	\label{rayleigh_eigen}
\end{equation}
where
\begin{equation}
	\tilde{I}_0 = I_0^{e,num} (\Psi_{\vec{k}}^*,\Psi_{\vec{k}}),
	\hspace{1em}
	\tilde{I}_{\alpha\beta,Z} = I_{\alpha\beta,Z}^{e,num} (\Psi_{\vec{k}}^*,\Psi_{\vec{k}}).
		\label{rayleigh_mat}
\end{equation}
Solving eq.~(\ref{rayleigh_eigen}), we obtain the frequencies of quasi-P and S-waves for numerical operators
(we define the larger one as the quasi-P-wave frequency).
We evaluate the numerical dispersion for P and S-waves as follows:
\begin{equation}
   \mbox{Dispersion}= \frac{
   \left|\omega_{\vec{k}}-V_\gamma |\vec{k}|\right|}{V_\gamma |\vec{k}|}
   \times 100\,\%,
   \label{e0}
\end{equation}
where $V_\gamma$ represents the P-wave velocity $V_P=\sqrt{(\lambda+2\mu)/\rho}$ or
the S-wave velocity $V_S=\sqrt{\mu/\rho}$.

Here we define $\rho=5$~$\mathrm{g}/\mathrm{cm}^3$,
$V_P=10$~$\mathrm{km}/\mathrm{s}$ and
$V_S=5$~$\mathrm{km}/\mathrm{s}$.
Fig.~\ref{fig0} shows the numerical dispersion properties of
our method and the SEM of several orders.
The errors
are plotted as functions of the average number of grid points per wavelength $G$.
We define $G$ as follows:
\begin{equation}
\nonumber
G=\frac{\ell}{h/n}=\frac{2n\pi}{h|\vec{k}|},
\label{e-1}
\end{equation}
where $\ell$ and $|\vec{k}|$ denote the wavelength and wavenumber
of a P or S-wave,
and $(h/n)$ indicates an average of the length between two adjacent grid intervals
along the $x$ and $y$-axes.
We see from Fig.~\ref{fig0} that our method has
the dispersion error smaller than the SEM of the same order
and is comparable to the SEM of one order higher,
which follows our theoretical results in Section~\ref{sec.theory_assmpt}.
Note that we use the mass matrix of Eq.~(\ref{66}) for the analysis of our method
(see \ref{appendix_c}).

Next, we discuss effects of temporal discretization on the numerical dispersion.
In this paper, we use the second-order finite-difference temporal operator.
In this case,
the numerical frequencies are replaced as follows \cite{SO08a, SO08b}:
\begin{equation}
	\omega_{\vec{k},\Delta t} =2\sin^{-1}(\omega_{\vec{k}} \Delta t/2) / \Delta t,
	\label{rayleigh_eigen_delt}
\end{equation}
where $\Delta t$ is the time interval,
$\omega_{\vec{k},\Delta t}$ is the numerical frequency for the time-discrete case,
and $\omega_{\vec{k}}$ is obtained by Eq.~(\ref{rayleigh_eigen}).
We define $\Delta t =CFL (h/n) / V_P$ with the CFL-number $CFL$.
Fig.~\ref{fig_rev1} shows the error of the frequencies of Eq.~(\ref{rayleigh_eigen_delt})
for several values of $CFL$.
We show for the propagation angle $\theta=\tan^{-1}(7/4)$,
which corresponds to the direction of the vector
from the source to the receiver point of Fig.~\ref{fig1}
used in Section~\ref{subsec.comp_wave}.

The numerical dispersion is significantly affected by the temporal discretization when $n>1$.
This is because the temporal operator has the dispersion error of $O(\Delta t^2)$,
and then, noting $\Delta t \propto h$, it is translated into $O(h^2)$,
regardless of smaller contributions of the spatial operators.
The downward peaks in the P-wave cases of Fig.~\ref{fig_rev1}
are due to sign reversal of the difference $(\omega_{\vec{k},\Delta t}-V_\gamma|\vec{k}|)$
in Eq.~(\ref{e0}).
For the P-wave cases,
effects of the temporal discretization clearly appear even at smaller values of $G$.
However, in general,
numerical dispersion of S-waves will be more considerably reflected in accuracy of wave computation,
since supposing that P and S-waves are excited by a unique source with a given frequency,
the S-wavelength (and the value of $G$)
is always $V_P/V_S$ times smaller than the P-wavelength.
\clearpage
\begin{figure}
\begin{center}
\includegraphics[width=30em]{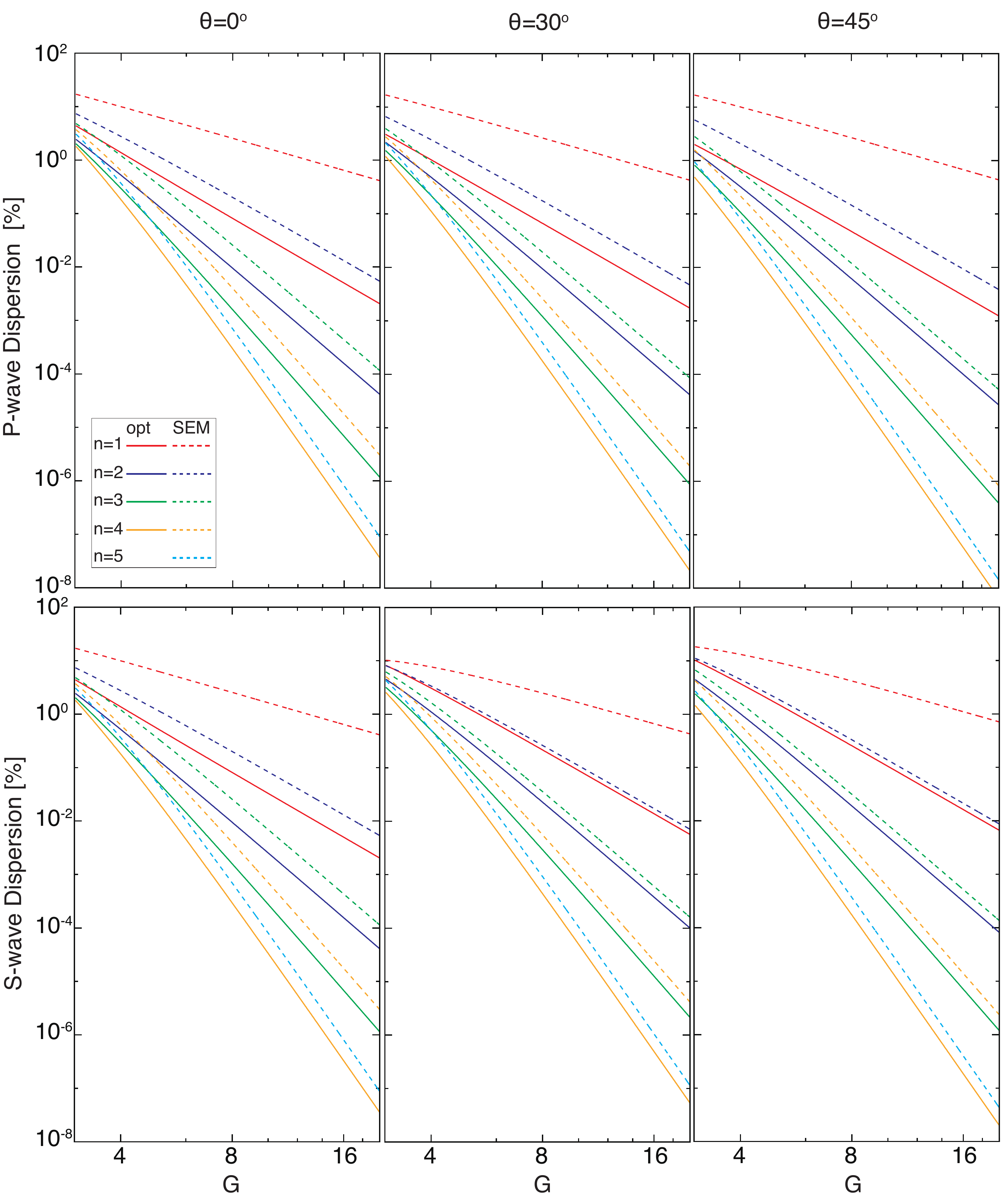}
\end{center}
\caption{
Numerical dispersion versus the average number of grid points per wavelength $G$.
$\theta$ denotes the angle between the wavenumber vector $\vec{k}$
and the positive $x$-axis.
Left, center, and right panels show for $\theta=0^\circ$, $30^\circ$, and $45^\circ$,
respectively,
and top and bottom panels for P and S-waves.
The solid and dashed lines correspond to our method and the SEM,
respectively.
For each case, colors red, blue, green, and orange correspond
to orders $n=1$, 2, 3, and 4, respectively,
and cyan corresponds to the SEM of order $n=5$.
(For interpretation of the references to color in this figure legend,
the reader is referred to the web version of this article.)
}
\label{fig0}
\end{figure}

\begin{figure}
\begin{center}
\includegraphics[width=30em]{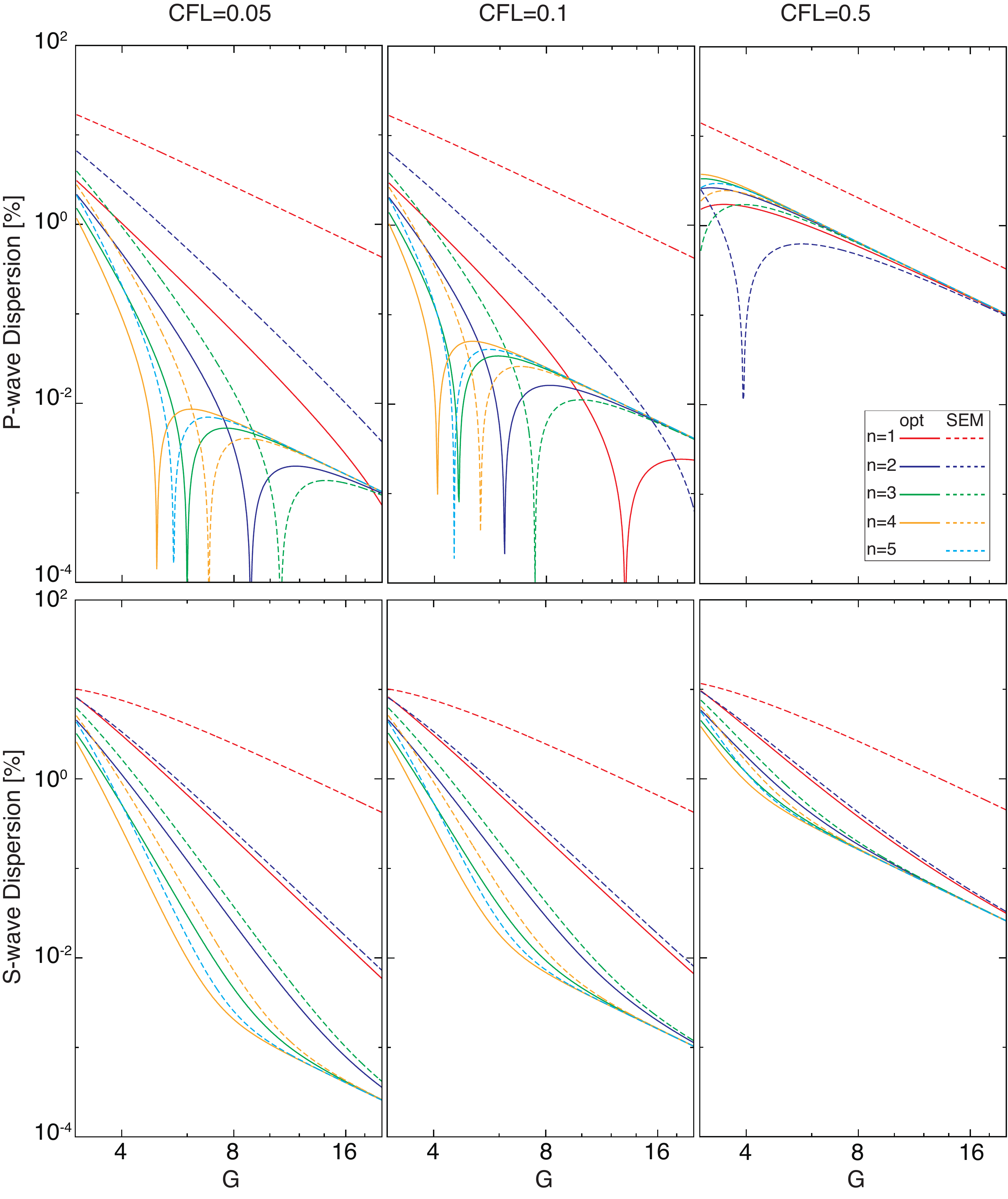}
\end{center}
\caption{
Effects of the second-order finite-difference temporal operator
on the numerical dispersion.
Left, center, and right panels show for $CFL=0.05$, $0.1$, and $0.5$,
respectively.
For each case, we use $\theta=\tan^{-1}(7/4)$.
The line types and colors represent the same as Fig.~\ref{fig0}.
}
\label{fig_rev1}
\end{figure}
\clearpage

\subsection{Waveform computation}\label{subsec.comp_wave}
We give a numerical comparison study between our method and the SEM
for the solutions
of the time-domain isotropic elastic wave equation for homogeneous and heterogeneous models,
in order to show that our method effectively improves the numerical solution.
Herein, we consider four cases of model structures: (a) homogeneous and (b)--(d) vertically ($y$-directionally) heterogeneous models.
Each model is a square of 2-D isotropic elastic medium
surrounded by free-surface boundaries with
the size 10~km $\times$ 10~km.
As shown in Fig.~\ref{fig1}, a source and receiver are located at $\vec{x}_s=(3\mbox{ km}, 3\mbox{ km})$
and $\vec{x}_r=(7\mbox{ km}, 10\mbox{ km})$, respectively,
with the left-bottom corner of a square as the origin point $(x,y)=(0\mbox{ km}, 0\mbox{ km})$.
Figs.~\ref{fig2}a--d show the vertical profiles of the P and S-velocity structures for models (a)--(d), respectively.
For the source, we use a point single force with the Ricker-wavelet as the source time function:
\begin{equation}
   f_\alpha (t, \vec{x})=
   g_\alpha (
   2\pi^2 \nu_c^2 t^2
   -1)\exp(- \pi^2 \nu_c^2 t^2)
    \delta(\vec{x}-\vec{x}_s),
    \label{e1}
\end{equation}
where
$\alpha$ represents $x$ and $y$-axes,
$g_\alpha$ are constants, and $\nu_c$ is the central frequency.
Here, we use
$(g_x,g_y)=(1\mbox{ N},1\mbox{ N})$ and $\nu_c=25$~$\mathrm{Hz}$.

For discretization,
we use the Cartesian grid with square elements of lengths $\Delta x=\Delta y=h$,
where cases of several values of $h$ are considered.
We set $CFL=0.05$ for the time interval.
The value of $\Delta t$ used in this section may be rather strict compared with
that used in actual applications of the SEM \cite{KV98,KT99}, in order to remove out the error
due to the temporal discretization as much as possible,
on which we do not focus in this analysis.
We use our modified operators
for the numerical spatial integration for each element except for the elements in contact with the bottom and left boundaries,
along which we cannot implement the modified operator~(\ref{51}) by the definition.
Thus, we use the ordinary SEM operators only for these elements.
For the numerical temporal integration,
we use the second-order finite-difference operator.
Note that although the mass matrix based on our method
is no longer diagonal,
we can avoid computing the inverse mass matrix
by using a predictor-corrector scheme for a non-diagonal mass matrix \cite{GMH12}
(see \ref{appendix_c} for detailed description).
The temporal integration is considered for the time
from $t_0=-80$~$\mathrm{ms}$ to $t_1=2920$~$\mathrm{ms}$.
For comparison,
we also make the same computation using the ordinary SEM operators.

Firstly, we consider model (a):
a homogeneous model with $\rho=5$~$\mathrm{g}/\mathrm{cm}^3$, $V_P=10$~$\mathrm{km}/\mathrm{s}$, and $V_S=5$~$\mathrm{km}/\mathrm{s}$,
which are exactly same as those used in the dispersion analysis of Section~\ref{subsec.numer_disp}.
Fig.~\ref{fig3}a shows the error of the waveforms at the receiver,
computed by using our method and the SEM of several orders.
The errors are plotted as functions of the average number of grid points per S-wavelength $G_c$.
We define $G_c$ as follows:
\begin{equation}
\nonumber
G_c=\frac{\ell_c}{h/n}=\frac{nV_S}{h\nu_c},
\label{e2}
\end{equation}
where $\ell_c$ denotes the length of the S-wavelet excited by the source~(\ref{e1}),
and we assume that $\ell_c$ is simply defined as the wavelength of the harmonic S-wave with frequency $\nu_c$:
i.e.,  $\ell_c = V_S /\nu_c$.
We evaluate the error of a waveform as follows:
\begin{equation}
   \mbox{Waveform Error}= \sqrt{\frac{\int_{t_0}^{t_1}
   \left|\vec{u}^{\,num}(t,\vec{x}_r)-\vec{u}(t,\vec{x}_r)\right|^2dt}{
   \int_{t_0}^{t_1}
   \left|\vec{u}(t,\vec{x}_r)\right|^2dt}}
   \times 100\,\%,
   \label{e3}
\end{equation}
where $\vec{u}^{\,num}$
is the displacement computed numerically,
and $\vec{u}$ is a reference solution
that is computed by using the SEM with higher-order elements
and an extremely fine grid ($n=5$ and $G_c=40$).

We see from Fig.~\ref{fig3}a that our method
has higher accuracy than the SEM of the same order,
and is roughly comparable to the SEM of one order higher.
Moreover, fortunately,
the curves for our method of orders $n=3$ and $4$ show superior convergence rates
of the accuracy even comparable to those of the SEM of two orders higher, respectively.
On the other hand,
the accuracy of our method of $n \leq 2$
seems to become slightly inferior to the SEM of one order higher with larger values of $G_c$.
This may be due to the errors of
the source representation of the ordinary SEM \cite{KT99,I16},
which we use in both of the methods:
i.e., the source term error will travel on the wavefield to the receiver,
and then can degrade the accuracy compared to the SEM of one order higher,
when $G_c$ becomes large such that the force term error dominates the dispersion error
in the overall accuracy.
Furthermore, since the $\delta$-point function (\ref{e1}) contains high wavenumber components,
lower-order ($n \leq 2$) source representations
can severely contaminate waveforms \cite{I16}.
When $n\geq 3$,
the accuracy seems to be somewhat plateauing when the error is smaller than 1~\%.
In those cases, the error due to the numerical temporal integration
is no longer negligible despite the small time interval
as shown in Fig.~\ref{fig_rev1}.
Fig.~\ref{fig4} shows a comparison of the error of waveforms computed by using our method and the SEM. Both for the cases, we use $n=2$ and $G_c=16$.
In this case, the errors of our method and the SEM are 2.4 and 19.1~\%, respectively.
Thus, the error of our method is roughly 8 times smaller than that of the SEM.
The CPU-time required for computation using our method
is roughly 1.4-1.7 times larger than that for the SEM of the same order,
regardless of the number and order of elements.

Next, we consider heterogeneous cases.
To simply characterize scales of heterogeneity of model structures,
we consider trigonometric functions with several different periods
for heterogeneous models, as shown in Figs.~\ref{fig2}b--d.
For models (b)--(d)
the lengths of periods are given by 4, 2, and 1~$\mathrm{km}$, respectively,
and the amplitudes are
20~\% from the standard P and S-velocities $V_P=10\mathrm{ km}/\mathrm{s}$ and
$V_S=5\mathrm{ km}/\mathrm{s}$.
The density $\rho$ and the other settings including 
the value of $\Delta t$ and the definition of $G_c$ are same as the case of model (a).
Figs.~\ref{fig3}b--d show
the error of the waveforms for models (b)--(d)
computed by using our method and the SEM of several orders.
We see similar trends in the case of model (a).
Note that even though
the accuracy of our method meets slight degradation from the homogeneous case,
especially when $n=3$ and $4$,
it still is comparable to the SEM of one higher order.
Since we use the same codes as the case of model (a),
so are the CPU-times for these cases.
\clearpage
\begin{figure}
\begin{center}
\includegraphics[width=15em]{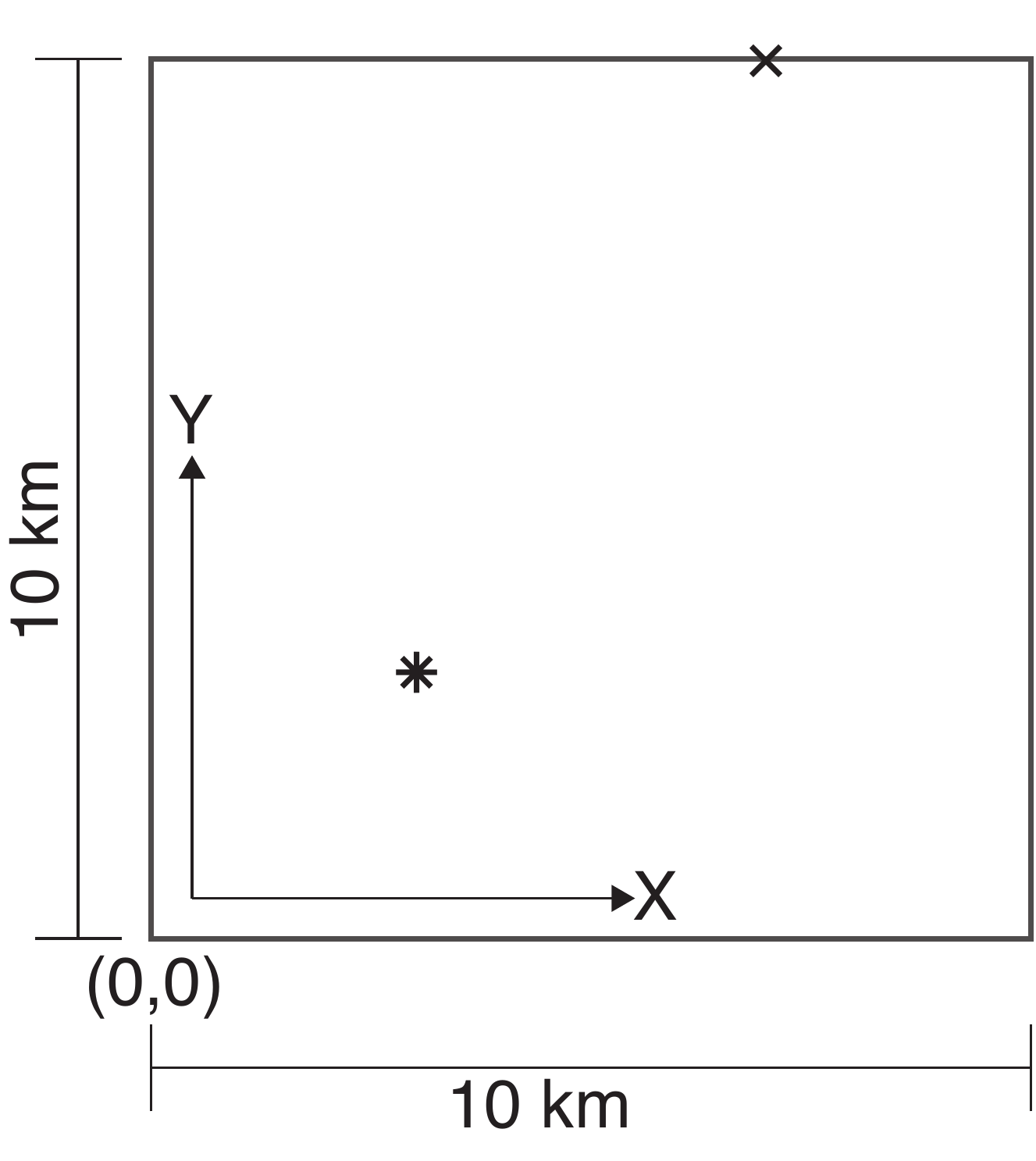}
\end{center}
\caption{
A schematic illustration of a 2-D model used in the waveform computations
of Section~\ref{sec.example}.
The star and cross indicate
the locations of source and receiver points, respectively.
}
\label{fig1}
\end{figure}
\clearpage
\begin{figure}
\begin{center}
\includegraphics[width=40em]{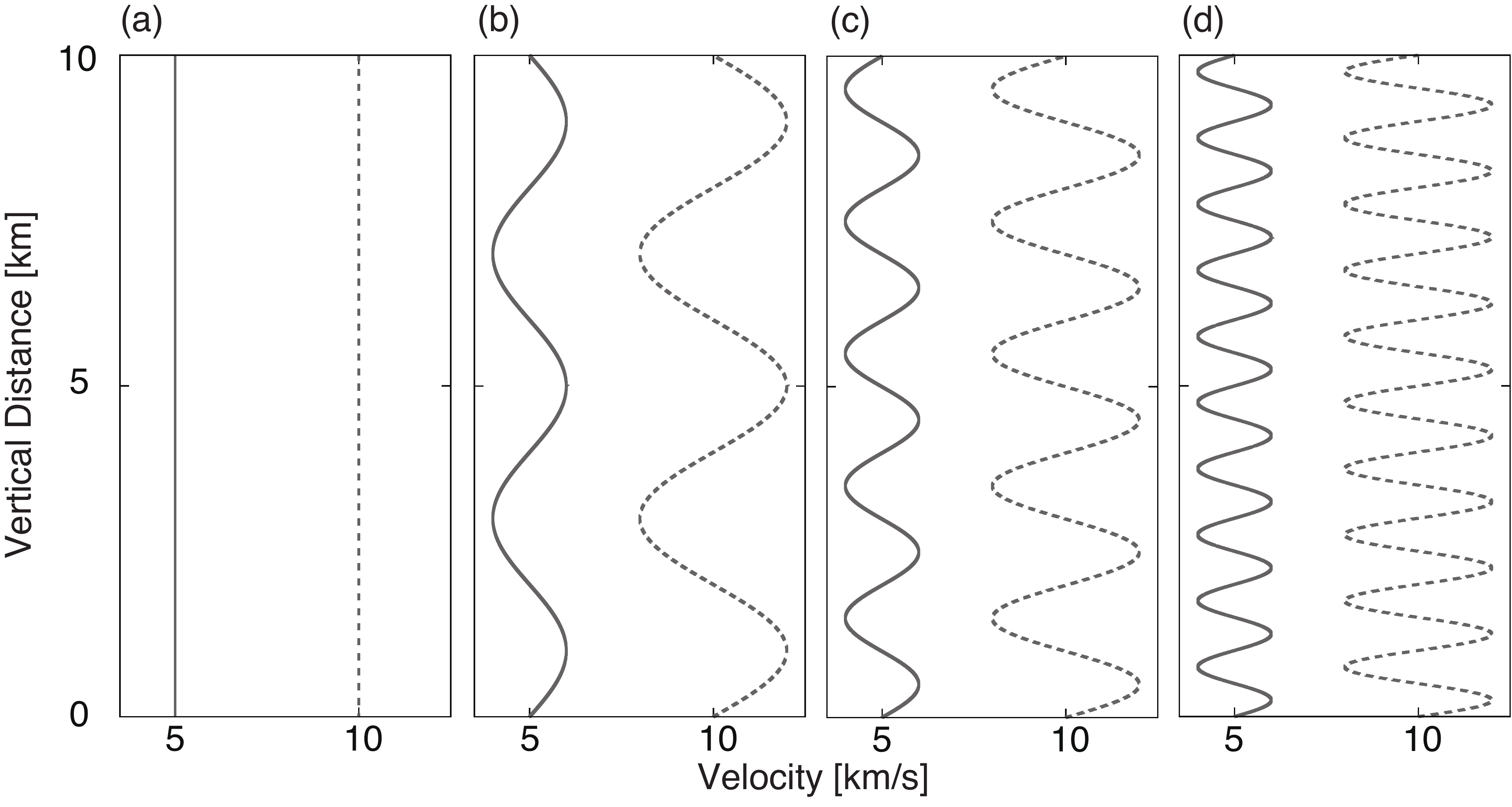}
\end{center}
\caption{
The vertical (the $y$-directional) dependences of models (a)--(d) used in the waveform computations of Section~\ref{sec.example}.
The dashed and solid lines show the P and S-wave velocities, respectively.
}
\label{fig2}
\end{figure}
\begin{figure}
\begin{center}
\includegraphics[width=40em]{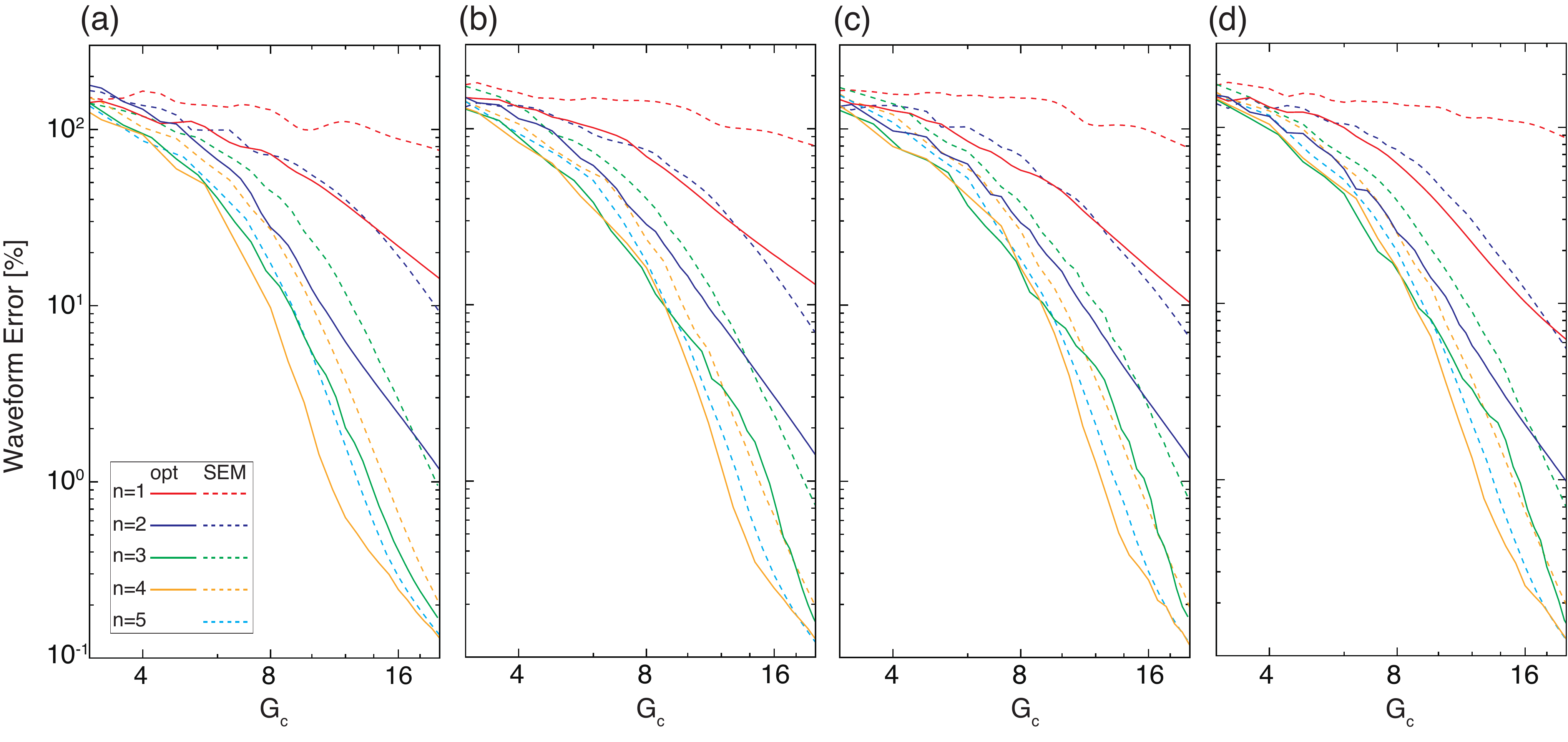}
\end{center}
\caption{
The error of computed waveforms versus the average number of grid points per S-wavelength $G_c$.
The line types and colors represent the same as Fig.~\ref{fig0}.
Panels (a)--(d) correspond to models (a)--(d) of Fig.~\ref{fig2}, respectively.}
\label{fig3}
\end{figure}
\begin{figure}
\begin{center}
\includegraphics[width=20em]{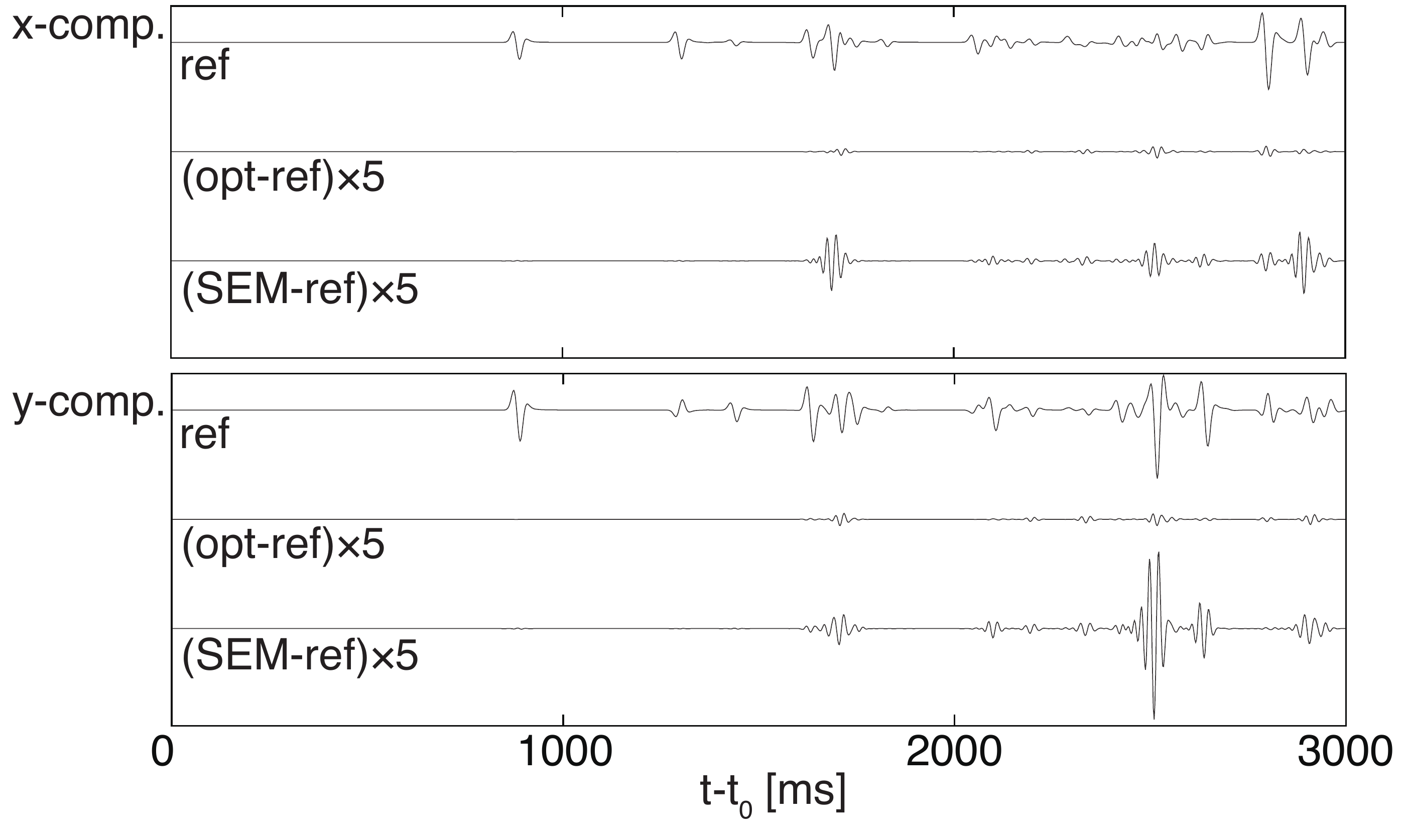}
\end{center}
\caption{
Waveform errors.
(top panel)
The first trace shows the $x$-component of the reference waveform used for model (a).
The second trace shows
the $x$-component of the residual of a waveform computed by our method
minus the reference solution,
and the third trace shows the residual for the SEM (both of the traces are amplified by 5 times).
We show for $n=2$ and $G_c=16$.
(bottom panel) The same as the top panel, except for the $y$-component.
}
\label{fig4}
\end{figure}
\clearpage
\section{Conclusions and future outlook}\label{sec.disc}
We derived modified numerical integration operators for the SEM
for simulation of elastic wave propagation.
In Section~\ref{sec.3-Dops},
we used these operators to define mass and stiffness matrices.
Although we omitted the derivation,
taking into account of the error due to the Lagrange interpolations
and the GLL rule \cite{KS05, J2013},
the respective errors of the mass and stiffness matrices for homogeneous and heterogeneous cases
are estimated to $O(h^{2n})$, as same as the SEM.
Nevertheless,
the numerical dispersion of the modified SEM
is suppressed to $O(h^{2n+2})$,
since the modification cancels the lower-order contributions of the respective
operators, as shown in Section~\ref{sec.theory_assmpt}.
We also showed numerical dispersion analysis and examples of computation of waveforms using our operators in order to follow our theoretical results. 
Although further comprehensive error analysis would be expected,
the numerical examples in Section~\ref{subsec.comp_wave}
show that our method effectively improves the accuracy of computation
even for several heterogeneous models.
While the optimally blending operator~(\ref{47}) itself has been previously presented by
\cite{AW10} with a rather strict approach,
we derived it on the basis of a perturbation error analysis~\cite{GT95},
which simplifies the derivation.
Furthermore, we extended this approach to elastic wave computation,
by introducing a new operator (\ref{51}).

In Section~\ref{subsec.comp_wave}, for simplicity,
we used the ordinary second-order finite-difference operator for time discretization.
However, as shown in Section~\ref{subsec.numer_disp},
the second-order operator would not be suitable for taking full advantage of higher-order spatial operators,
because of its non-negligible contamination against the smaller errors of the spatial operators,
when we use a larger value of the CFL number.
In practical applications of our method for the order $n>1$ cases
(and also the original SEM for the order $n>2$ cases,
as pointed out by Oliveira and Seriani \cite{OS10}),
it would be more preferable to use a higher-order temporal operator such as the Lax--Wendroff method \cite{LW64,MGT00}.

Whereas our method is derived based on the Cartesian grid,
a deformed grid would more flexibly meet configurations of surface and internal boundaries.
Nevertheless, considering the mapping onto a reference coordinate system,
our method can still be implemented as tensor products of the
integration operators defined for the reference coordinate axes, as in the SEM
\cite{KS05}.
However, in this case,
effects of grid deformation are incorporated
as heterogeneity of physical parameters on the reference coordinate system,
even when the original model is homogeneous.
Moreover, our method is limited to structured grids,
since the integration (\ref{51}) uses a node outside the domain of integration,
and then all quadrature nodes should be aligned on a single line. 
Although, unfortunately, it might be difficult to estimate effects of grid deformation
based on the theoretical framework proposed in this study,
these effects on the numerical dispersion can be estimated based on a numerical approach \cite{OS11}.
This might be a starting point for error analysis of our method with grid deformation.
This extension and error analysis yet should be important subjects for future work.

\section*{Acknowledgements}
We thank the two anonymous reviewers for carefully reading our manuscript
and giving helpful comments.
This research was partly supported by grants from
the Ministry of Science and Technology of Taiwan
(MOST-106-2811-M-001-158).
N. F. is partly supported by GPX consortium of
Institut de Physique du Globe de Paris,
\'{E}cole des Mines de Paris,
Schlumberger, CGG, TOTAL, and
Agence Nationale de la Recherche of France (ANR-12-CHIN-0003).

\appendix
\section{Basic formulas}\label{appendix_a}
In the definition of the numerical integration operators
and their error analysis in Section~\ref{sec.theory_assmpt},
we often use the following lemma:
\begin{equation}
 \int_{-1}^1 P_n (x) x^l dx=
 \sum_{i=0}^n q_i P_n(x_i) x_i^l 
  =0\hspace{1em}\mbox{if}\hspace{0.5em}l<n,
     \label{a10}
\end{equation}
where $l$ is a non-negative integer.
This relation is obtained by using the following orthogonality relation \cite{J2013}:
\begin{equation}
   \int_{-1}^1 P_{n'}(x) P_n(x)dx = \frac{2}{2n+1}\delta_{n'n},
        \label{a2}
\end{equation}
and recalling that the GLL quadrature rule exactly computes an integral
when the integrand is a polynomial of degree $(2n-1)$ or below.

The GLL nodes $x_i$ are the zeros of the following polynomial \cite{KS05,I16}:
\begin{equation}
  W(x) \equiv (x^2-1)P'_n(x)
          = (n+1)\left[P_{n+1}(x)-xP_n(x)\right].
   \label{a6}
\end{equation}
The GLL weights are given by
\begin{equation}
   q_i = \int_{-1}^1 L_i(x) dx
   =\frac{2}{n(n+1) \left[P_n(x_i)\right]^2},
      \label{a8}
\end{equation}
with $i=0,\dots,n$.

A polynomial of degree $n$ or below
can be exactly expressed in terms of a polynomial basis of degree $n$.
Then, the constant $1$ can be written by a linear combination of $L_i$ with $i=0,\dots,n$ \cite{J2013},
as follows:
\begin{equation}
  1= \sum_{i=0}^n L_i(x).
        \label{a11}
\end{equation}
Taking the differentiation of Eq.~(\ref{a11}),
we have
\begin{equation}
  \sum_{i=0}^n L'_i(x) = 0.
          \label{a12}
\end{equation}

In addition, we list
some notable formulas.
The following formulas are well-known \cite{KS05,I16,J2013},
or immediately obtained by using well-known formulas:
 \begin{eqnarray}
    L_i(x_j) &=& \delta_{ij}
                   \label{a17}\\
     L_i'(x_j) &=& \frac{P_n'(x_i)}{2P_n(x_i)} \delta_{ij} +(1-\delta_{ij}) \frac{P_n(x_j)}{P_n(x_i)}\frac{1}{x_j-x_i}
    \label{a21}\\
   W'(x) &=&[(x^2-1)P_n'(x)]'
   =n(n+1)P_n(x)
                \label{a15}\\
  \int_{-1}^1 1dx&=&\sum_{i=0}^n q_i =2
     \label{a9}\\
   L_i(x) &=& \frac{1}{W'(x_i)}\frac{W(x)}{x-x_i}
   =\frac{1}{n+1}\frac{P_n(x)}{P_n(x_i)}+o(x^{n-1})
                   \label{a18}\\
    \int_{-1}^1 P_n (x) x^n dx &=&\frac{2}{2n+1}a_n^{-1}
     \label{a5}\\
   \frac{L_i(x)-L_i(x_j)}{x-x_j}
   &=&  \frac{na_n}{W'(x_i)} x^{n-1} + o(x^{n-2}),
                  \label{a19}
\end{eqnarray}
where $i, j=0,\dots,n$,
$o(x^l)$ represents terms of degree $l$ or below,
and $a_n$ represents the coefficient of the $x^n$-term of polynomial $P_n(x)$.
Note that $o(x^l)$ is not to be confused with
$O(x^l)$, where $O(x^l)$ represents terms of degree $l$ or above.

By the definition~(\ref{49}), the following formulas are obtained for $X_i$
\begin{eqnarray}
   X'_i(x)&=&\left\{
   \begin{array}{cl}
   \displaystyle{L'_i(x)+\frac{P_n(x)}{(x_i-x_{-1})P_n(x_i)}} &
   \mbox{if}\hspace{0.5em}  i \neq -1\\
   \vspace{-1em}\\
    \displaystyle{\frac{n(n+1)P_n(x)}{(x_{-1}^2-1)P_n'(x_{-1})}} &
       \mbox{if}\hspace{0.5em}  i = -1
   \end{array}\right.
             \label{a22}\\
       \sum_{i=-1}^n X'_i(x) &=& 0
          \label{a13}\\
   X_i(x_j) &=& \delta_{ij},
       \label{a_extra1}
\end{eqnarray}
where $i,j=-1,0,\dots,n$, and Eq.~(\ref{a13}) is derived from the same reason as Eq.~(\ref{a12}).

Finally, we note the following integration results:
\begin{eqnarray}
   \nonumber
    \int_{-1}^1 L_i(x)L_j(x) dx
   &=& \frac{1}{W'(x_i)}\int_{-1}^1\frac{W(x)}{x-x_i} L_j(x) dx\\
      \nonumber
   &=& \frac{L_j(x_i)}{W'(x_i)}\int_{-1}^1\frac{W(x)}{x-x_i} dx
   + \frac{1}{W'(x_i)}\int_{-1}^1W(x) \frac{L_j(x)-L_j(x_i)}{x-x_i} dx\\
   &=& \delta_{ij}q_i- \frac{n(n+1)}{2(2n+1)}q_iP_n(x_i) q_jP_n(x_j)
       \label{a23}\\
         \nonumber
  \int_{-1}^1 L_i(x) P_n(x) dx&=&
\frac{1}{(n+1)P_n(x_i)}\int_{-1}^1 \left[P_n(x)+o(x^{n-1})\right] P_n(x) dx\\
  &=&\frac{n}{2n+1}q_iP_n(x_i)
         \label{a24}\\
  \int_{-1}^1 X'_{i>-1}(x) L_j(x) dx
  &=&
  \int_{-1}^1 L'_i(x)L_j(x)dx
  +\frac{n^2(n+1)}{2(2n+1)}\frac{q_iP_n(x_i) q_j P_n(x_j)}{x_i-x_{-1}}
           \label{a25}\\
  \int_{-1}^1 X'_{-1}(x) L_j(x) dx
  &=&
  \frac{n^2(n+1)}{2n+1}\frac{q_jP_n(x_j)}{(x^2_{-1}-1)P'_n(x_{-1})}.
         \label{a26}
\end{eqnarray}
\section{Error-estimation of numerical integration operators}\label{appendix_b}
Herein, we derive Eqs.~(\ref{40})--(\ref{42}), (\ref{46}), and (\ref{52})--(\ref{54}).
For these purposes,
we substitute $(\phi,\psi)=(p_{\hat{h}}^*,p_{\hat{h}})$ in the integration operators
(\ref{33})--(\ref{35}), (\ref{45}), (\ref{47}), and (\ref{51}),
and thereby immediately obtain their relative errors by comparing them with the exact values given by Eqs.~(\ref{37})--(\ref{39}).
$A^{SEM}(p_{\hat{h}}^*, p_{\hat{h}})$
is immediately obtained as follows:
\begin{equation}
   A^{SEM}(p_{\hat{h}}^*, p_{\hat{h}}) = \sum_{i=0}^n q_i =2,
   \label{b1}
\end{equation}
where we use Eq.~(\ref{a9}).
$A^{FEM}(p_{\hat{h}}^*, p_{\hat{h}})$ is 
calculated as follows:
\begin{eqnarray}
   \nonumber
    A^{FEM}( p_{\hat{h}}^*, p_{\hat{h}})
   &=& 2
   -\frac{n(n+1)}{2(2n+1)}\left|\sum_{i=0}^nq_iP_n(x_i) \mathrm{e}^{\mathrm{i} \hat{h} x_i} \right|^2 \\
   \nonumber
 &=& 2 -\frac{n(n+1)}{2(2n+1)}\left|\sum_{i=0}^nq_iP_n(x_i)\sum_{l=0}^\infty \frac{(\mathrm{i} \hat{h} x_i)^l}{l!}\right|^2\\
    \nonumber
   &=& 2 -\frac{n(n+1)\hat{h}^{2n}}{2(2n+1)(n!)^2} \left[\sum_{i=0}^nq_iP_n(x_i)x_i^n \right]^2 +O(\hat{h}^{2n+2})\\
   &=& 2 +2(n+1)\mathcal{F}_n \hat{h}^{2n}+O(\hat{h}^{2n+2}),
      \label{b2}
\end{eqnarray}
where $\mathcal{F}_n$ is given by Eq.~(\ref{43}),
and we use Eq.~(\ref{a10}) for the derivation of the third line.
Then, using the first line of Eq.~(\ref{47}),
and Eqs.~(\ref{b1}) and (\ref{b2}),
we immediately obtain
 \begin{equation}
    A^{opt}( p_{\hat{h}}^*, p_{\hat{h}})
    = 2+2\mathcal{F}_n \hat{h}^{2n}+O(\hat{h}^{2n+2}).
          \label{b3}
 \end{equation}
 
In order to calculate
$B^{SEM}(p_{\hat{h}}^*, p_{\hat{h}})$, let us first evaluate:
\begin{eqnarray}
       \nonumber
       S_j&\equiv&
\sum_{i=0}^n \mathrm{e}^{\mathrm{i} \hat{h} x_i} L'_i(x_j)\\
       \nonumber
&=& \mathrm{e}^{\mathrm{i} \hat{h} x_j} \sum_{i=0, i\neq j}^n
\left[ \mathrm{e}^{\mathrm{i} \hat{h} (x_i-x_j)} - 1\right] L'_i(x_j)\\
       \nonumber
&=& -\mathrm{e}^{\mathrm{i} \hat{h} x_j} P_n(x_j) \sum_{i=0, i\neq j}^n
 \frac{1}{P_n(x_i)} \frac{\mathrm{e}^{\mathrm{i} \hat{h} (x_i-x_j)} - 1}{x_i-x_j}\\
        \nonumber
 &=& -\mathrm{e}^{\mathrm{i} \hat{h} x_j} P_n(x_j) \sum_{i=0}^n
 \frac{(1-\delta_{ij})}{P_n(x_i)} \sum_{l=1}^\infty \frac{(\mathrm{i} \hat{h})^l(x_i-x_j)^{l-1}}{l!}\\
  &=& \left[ \mathrm{i}\hat{h}-\frac{n(n+1)}{2}P_n(x_j) R_j \right]\mathrm{e}^{\mathrm{i} \hat{h} x_j},
            \label{b4}
\end{eqnarray}
where $R_j$ in the last line is
\begin{equation}
   R_j 
 \equiv \sum_{i=0}^nq_iP_n(x_i) \sum_{l=1}^\infty \frac{(\mathrm{i} \hat{h})^l(x_i-x_j)^{l-1}}{l!}.
             \label{b5}
\end{equation}
Note that we use Eq.~(\ref{a12}) for the derivation of
the second line of Eq.~(\ref{b4}), and Eq.~(\ref{a21}) for the third line.
Further, we evaluate the following values:
\begin{eqnarray}
\nonumber
  E&\equiv&\sum_{j=0}^n| R_j|^2\\\nonumber
   &=&\sum_{j=0}^n\frac{\hat{h}^{2n+2}}{[(n+1)!]^2} \left[\sum_{i=0}^nq_iP_n(x_i) x_i^n\right]^2 + O(\hat{h}^{2n+4})\\
  &=&\frac{\hat{h}^{2n+2}}{(n+1)(n!)^2} \left[\sum_{i=0}^nq_iP_n(x_i) x_i^n\right]^2 + O(\hat{h}^{2n+4})
               \label{b6}\\
\nonumber
  T&\equiv&\sum_{j=0}^nq_jP_n(x_j) ( R_j- R^*_j)\\
  \nonumber
    &=&\sum_{i, j=0}^n
  q_jP_n(x_j) q_i P_n(x_i) \sum_{l=0}^\infty 
  \frac{2(\mathrm{i}\hat{h})^{2l+1}(x_i-x_j)^{2l} }{(2l+1)!}\\
   &=& \mathrm{i} \hat{h}^{2n+1}  \frac{2{{2n}\choose{n}}  }{(2n+1)!}\left[ \sum_{i=0}^n q_i P_n(x_i)x_i^n\right]^2+O(\hat{h}^{2n+3}),
                  \label{b7}
\end{eqnarray}
where we use Eq.~(\ref{a10}),
and ${{2n}\choose{n}}$ denotes the binomial coefficient of the $(-1)^n x_i^n x_j^n$-term
in the expansion of $(x_i-x_j)^{2n}$.
Finally, we obtain
\begin{eqnarray}
   \nonumber
   B^{SEM}( p_{\hat{h}}^*, p_{\hat{h}})&=&\sum_{j=0}^n q_j \left| S_j \right|^2\\
         \nonumber
   &=& \sum_{j=0}^n q_j \left|\mathrm{i}\hat{h}-\frac{n(n+1)}{2} P_n(x_j) R_j \right|^2\\
      \nonumber
    &=&
    2\hat{h}^2+\frac{n(n+1)}{2}E +\mathrm{i}\hat{h} \frac{n(n+1)}{2} T\\
 &=&
   2\hat{h}^2
   +2\mathcal{F}_n\hat{h}^{2n+2} +O(\hat{h}^{2n+4}).
                  \label{b8}
\end{eqnarray}

$C^{SEM}(p_{\hat{h}}^*, p_{\hat{h}})$ is calculated as follows:
\begin{eqnarray}
\nonumber
     C^{SEM}(p_{\hat{h}}^*, p_{\hat{h}})&=&
           \sum_{j=0}^n q_j  \mathrm{e}^{\mathrm{i} \hat{h}x_j} S_j^*\\
      \nonumber
      &=& \sum_{j=0}^n q_j
      \left[-\mathrm{i} \hat{h} - \frac{n(n+1)}{2}P_n(x_j) R_j^* \right]\\
            \nonumber
      &=& -2 \mathrm{i} \hat{h}  - \frac{n(n+1)}{2} \sum_{i, j=0}^n q_j P_n(x_j)q_iP_n(x_i)
      \sum_{l=1}^\infty \frac{(-\mathrm{i} \hat{h} )^l(x_i-x_j)^{l-1}}{l!}\\
                      \nonumber
          &=& -2 \mathrm{i} \hat{h} 
          - \frac{n(n+1)}{2} \sum_{i, j=0}^n q_j P_n(x_j) q_iP_n(x_i)
     \frac{(-\mathrm{i} \hat{h})^{2n+1}{{2n}\choose{n}} x_i^n (-x_j)^n }{(2n+1)!}+O(\hat{h}^{2n+2})\\
                         \nonumber
           &=& -2 \mathrm{i} \hat{h} + \mathrm{i}\hat{h}^{2n+1}
         \frac{n(n+1) {{2n}\choose{n}} }{2(2n+1)!}      
           \left[\sum_{i=0}^n q_iP_n(x_i) x_i^n \right]^2+O(\hat{h}^{2n+2})\\
           &=& -2 \mathrm{i} \hat{h}  -2\mathrm{i}(n+1)\mathcal{F}_n\hat{h}^{2n+1}  +O(\hat{h}^{2n+2}),
                             \label{b9}
\end{eqnarray}
where we use Eq.~(\ref{a10}) for the derivation of the fourth line.

To calculate $C^{opt}(p_{\hat{h}}^*, p_{\hat{h}})$,
let us first evaluate:
\begin{eqnarray}
\nonumber
 Q(x) &\equiv& 
   \sum_{i=-1}^n \mathrm{e}^{\mathrm{i} \hat{h} x_i}X_i'(x)\\
   \nonumber
   &=&  \mathrm{e}^{\mathrm{i} \hat{h} x_{-1}}
   \sum_{i=-1}^n \left[\mathrm{e}^{\mathrm{i} \hat{h} (x_i-x_{-1})} -1\right]X_i'(x)\\
   \nonumber
   &=&  \mathrm{e}^{\mathrm{i} \hat{h} x_{-1}}
   \sum_{i=0}^n \left[\mathrm{e}^{\mathrm{i} \hat{h} (x_i-x_{-1})} -1\right]X_i'(x)\\
   \nonumber
   &= &
   \sum_{i=0}^n \mathrm{e}^{\mathrm{i} \hat{h} x_i}L_i'(x)
   +  
    \mathrm{e}^{\mathrm{i} \hat{h} x_{-1}}
   \sum_{i=0}^n\frac{P_n(x)}{P_n(x_i)} \frac{\mathrm{e}^{\mathrm{i} \hat{h} (x_i-x_{-1})} -1}{x_i-x_{-1}}\\
   &=& 
   \sum_{i=0}^n \mathrm{e}^{\mathrm{i} \hat{h} x_i}L_i'(x)+  
    \mathrm{e}^{\mathrm{i} \hat{h} x_{-1}} HP_n(x),
                                 \label{b10}
\end{eqnarray}
where $H$ in the last line is
\begin{eqnarray}
   \nonumber
  H&\equiv& \sum_{i=0}^n \frac{1}{P_n(x_i)}
  \frac{ \mathrm{e}^{\mathrm{i} \hat{h} (x_i-x_{-1})} -1}{x_i-x_{-1}}\\
  \nonumber
  &=&
    \frac{n(n+1)}{2}
   \sum_{i=0}^n q_i P_n(x_i)
   \sum_{l=1}^\infty \frac{(\mathrm{i} \hat{h})^l (x_i-x_{-1})^{l-1}}{l!}\\
   &=&
  (\mathrm{i} \hat{h})^{n+1}\frac{n}{2n!}
   \sum_{i=0}^n q_i P_n(x_i)x_i^n+O(\hat{h}^{n+2}).
                             \label{b11}
\end{eqnarray}
Note that we use 
Eq.~(\ref{a13}) for the derivation of the second line of Eq.~(\ref{b10}), Eq.~(\ref{a22}) for the fourth line of Eq.~(\ref{b10}),
and Eq.~(\ref{a10}) for the third line of Eq.~(\ref{b11}).
Finally, $C^{opt}(p_{\hat{h}}^*, p_{\hat{h}})$ is calculated as follows:
\begin{eqnarray}
\nonumber
   C^{opt}(p_{\hat{h}}^*, p_{\hat{h}})
   &=& \int_{-1}^1 \left[Q(x)\right]^* \sum_{j=0}^n \mathrm{e}^{\mathrm{i} \hat{h} x_j} L_j(x) dx\\
   \nonumber
  &=&
 \sum_{i,j=0}^n \mathrm{e}^{\mathrm{i} \hat{h} (x_j-x_i)}  \int_{-1}^1  L'_i(x) L_j(x) dx
 + H^* \sum_{j=0}^n \mathrm{e}^{\mathrm{i} \hat{h} (x_j-x_{-1})}
    \int_{-1}^1 L_j(x) P_n(x) dx\\
   \nonumber
   &=&
    C^{SEM}(p_{\hat{h}}^*, p_{\hat{h}})+
   \frac{n H^*}{2n+1}
   \sum_{j=0}^nq_jP_n(x_j) \mathrm{e}^{\mathrm{i} \hat{h} (x_j-x_{-1})}\\
   \nonumber
   &=&
    C^{SEM}(p_{\hat{h}}^*, p_{\hat{h}})+
   (\mathrm{i}\hat{h})^n \frac{n H^*}{(2n+1)n!}
   \sum_{j=0}^nq_jP_n(x_j)x_j^n
   + O({\hat{h}}^{n+1}) H^*\\
   \nonumber
   &=&
    C^{SEM}(p_{\hat{h}}^*, p_{\hat{h}})
    +2\mathrm{i}n \mathcal{F}_n\hat{h}^{2n+1}+O(\hat{h}^{2n+2})\\
   &=&
   -2\mathrm{i}\hat{h}-2\mathrm{i}\mathcal{F}_n\hat{h}^{2n+1}+O(\hat{h}^{2n+2}),
     \label{b12}
\end{eqnarray}
where we use
Eq.~(\ref{a24}) for the integration in the second term of the second line,
and Eq.~(\ref{a10}) for the derivation of the fourth line.
Note that although we here omit the evaluation of the coefficient of the
$\hat{h}^{2n+2}$ term in the last line of Eq.~(\ref{b12}),
we see that it is zero or a real number.
Hence, 
Eq.~(\ref{54})
has zero or a non-zero pure imaginary term at the $(2n + 1)$th-order.

\section{Predictor-corrector scheme for explicit time-marching}\label{appendix_c}
Matrix operator $\left(A_{ij}^{opt}\right)$ of Eq.~(\ref{3.5}) is no longer diagonal.
Then, supposing that the mass matrix is
defined on the basis of this matrix, as shown in
Eq.~(\ref{kuso4.2_9}) or (\ref{kuso4.3_9}),
it will not be consequently diagonal.
In that case, the inverse mass matrix would not be easily obtained.
However, we can avoid computing the inverse mass matrix by applying
a predictor-corrector time-marching scheme for a non-diagonal mass matrix \cite{GMH12}.
Here, we reformulate the scheme of \cite{GMH12} to be applicable to our method.

Supposing that
the mass and stiffness matrices
are constructed based on our modified operators,
the time-domain discrete form of the wave equation
can be written as follows:
\begin{equation}
   \mathbf{M}^{G}\ddot{\mathbf{U}}^t
   = -\mathbf{K}^{G}\mathbf{U}^t + \mathbf{F},
   \label{61}
\end{equation}
where $\mathbf{M}^{G}$ and $\mathbf{K}^{G}$ are the global mass and stiffness matrices
which are obtained by assembling the local mass and stiffness matrices
(\ref{kuso4.2_7}) and (\ref{kuso4.2_8}),
$\mathbf{U}^t$ and $\ddot{\mathbf{U}}^t$
are the discretized displacement and acceleration at the time $t$, respectively, and $\mathbf{F}$ is the force term.
Here, we suppose that
the acceleration $\ddot{\mathbf{U}}^t$ is approximated by the second-order finite-difference operator as follows:
\begin{equation}
   \ddot{\mathbf{U}}^t
   \approx \frac{\mathbf{U}^{t+\Delta t} -2\mathbf{U}^t + \mathbf{U}^{t-\Delta t}}{\Delta t ^2},
      \label{62}
\end{equation}
where $\mathbf{U}^{t+\Delta t}$ and $\mathbf{U}^{t-\Delta t}$
are the discretized displacements at the time $t+\Delta t$ and $t-\Delta t$, respectively.
Note that $\mathbf{U}^t$ and $\mathbf{U}^{t-\Delta t}$ are the variables already known,
whereas the components of $\mathbf{U}^{t+\Delta t}$
are unknown variables to be solved from Eqs.~(\ref{61}) and (\ref{62}).

Now we expand Eq.~(\ref{kuso4.2_9}) by using Eq.~(\ref{3.5}) as follows:
\begin{equation}
	A_{i_xj_x}^{opt}A_{i_yj_y}^{opt} =
	A_{i_xj_x}^{SEM}A_{i_yj_y}^{SEM}
	-\frac{n}{2(n+1)}\left(b_{i_x}b_{j_x} A_{i_yj_y}^{SEM}
	+
	 A_{i_xj_x}^{SEM}b_{i_y}b_{j_y}
	\right)+\frac{n^2}{4(n+1)^2}b_{i_x}b_{j_x}b_{i_y}b_{j_y}.
      \label{63}
\end{equation}
As shown in Section~\ref{sec.theory_assmpt},
both $A^{opt}$ and $A^{SEM}$ of Eqs.~(\ref{33}) and (\ref{47})
have the contributions of $O(\hat{h}^{2n})$ or above to the numerical dispersion.
Consequently, the difference between $A^{opt}$ and $A^{SEM}$
will have the contribution of $O(\hat{h}^{2n})$ or above.
Hence, the third term of the right-hand side of Eq.~(\ref{63})
will have contribution of only $O(\hat{h}^{4n})$ or above,
and thus its effect will be negligible.
Therefore, we decompose $\mathbf{M}^G$ as follows:
\begin{equation}
    \mathbf{M}^{G} \approx \mathbf{M}^{G}_{diag}
    +\mathbf{M}_{corr}^G,
      \label{66}
\end{equation}
where $\mathbf{M}^G_{diag}$ and $\mathbf{M}_{corr}^G$ are matrices which are
obtained by assembling the first and second terms of the
right-hand side of Eq.~(\ref{63}) defined for each element, respectively.
Note that $\mathbf{M}^G_{diag}$ is just the diagonal mass matrix of the SEM.

If we replace $\mathbf{M}^G$ of Eq.~(\ref{61}) by $\mathbf{M}^G_{diag}$,
we obtain the following equation:
\begin{equation}
   \mathbf{M}^G_{diag}\ddot{\mathbf{U}}^t_{pred} =
   -\mathbf{K}^G\mathbf{U}^t +\mathbf{F}
          \label{65}
\end{equation}
with a vector $\ddot{\mathbf{U}}^t_{pred}$.

We decompose $\ddot{\mathbf{U}}^t$ of Eq.~(\ref{61}) as follows:
\begin{equation}
    \ddot{\mathbf{U}}^t = \ddot{\mathbf{U}}^t_{pred}+\ddot{\mathbf{U}}^t_{corr},
      \label{67}
\end{equation}
where $\ddot{\mathbf{U}}^t_{corr}$ is the difference from the solution of Eq.~(\ref{65}).
Supposing that $\ddot{\mathbf{U}}^t_{corr}$ is sufficiently small so that
$\mathbf{M}_{corr}^G \ddot{\mathbf{U}}^t_{corr}$ is negligible, we have
\begin{equation}
   \mathbf{M}^G_{diag} \ddot{\mathbf{U}}^t_{corr}
   =- \mathbf{M}_{corr}^G \ddot{\mathbf{U}}_{pred}^t.
        \label{68}
\end{equation}
Since $\mathbf{M}^G_{diag}$ is diagonal, $\ddot{\mathbf{U}}^t_{corr}$
is immediately obtained from this equation.
We use Eqs.~(\ref{65})--(\ref{68}),
instead of Eq.~(\ref{61}), and then we obtain $\mathbf{U}^{t+\Delta t}$
for the next time step by using Eq.~(\ref{62}).
Note that the above scheme may be also applicable
to any other numerical temporal finite-difference operators, rather than Eq.~(\ref{62}).

\bibliography{mybibfile}

\end{document}